\documentclass[aps,prb,reprint,superscriptaddress,intlimits]{revtex4-2}
\usepackage{bm,latexsym,mathrsfs,enumerate,amsmath,amssymb}

\usepackage{graphicx}
\usepackage{ulem}
\usepackage[breaklinks=true,urlcolor = blue,colorlinks = true,citecolor = blue,linkcolor = blue]{hyperref}
\usepackage[utf8]{inputenc}

\begin{document}

\title{Nonequilibrium dynamics of magnetic hopfions driven by spin-orbit torque}

\author{Shoya Kasai}
  \email{Contact author: kasai@aion.t.u-tokyo.ac.jp}
  \affiliation{Department of Applied Physics, The University of Tokyo, Hongo, Tokyo 113-8656, Japan}
\author{Shun Okumura}
  \affiliation{Department of Applied Physics, The University of Tokyo, Hongo, Tokyo 113-8656, Japan}
  \affiliation{Quantum-Phase Electronics Center (QPEC), The University of Tokyo, Hongo, Tokyo 113-8656, Japan}
  \affiliation{RIKEN Center for Emergent Matter Science (CEMS), Wako, Saitama 351-0198, Japan}
\author{Yukitoshi Motome}
  \email{Contact author: motome@ap.t.u-tokyo.ac.jp}
  \affiliation{Department of Applied Physics, The University of Tokyo, Hongo, Tokyo 113-8656, Japan}

\begin{abstract}
Hopfions—three-dimensional topological solitons with knotted spin textures—have recently garnered attention in topological magnetism due to their unique topology characterized by the Hopf number $H$, a topological invariant derived from knot theory. In contrast to two-dimensional skyrmions, which are typically limited to small topological invariants, i.e., skyrmion numbers, hopfions can, in principle, be stabilized with arbitrary Hopf numbers. However, the nonequilibrium dynamics, especially interconversion between different Hopf numbers, remain poorly understood. Here, we theoretically investigate the nonequilibrium dynamics of hopfions with various Hopf numbers by numerically solving the Landau-Lifshitz-Gilbert equation with spin-orbit torque (SOT). For $H=1$, we show that SOT induces both translational and precessional motion, with dynamics sensitive to the initial orientation. For $H=2$, we find that intermediate SOT strengths can forcibly split the hopfion into two $H = 1$ hopfions. This behavior is explained by an effective tension picture, derived from the dynamics observed in the $H=1$ case. By comparing the splitting dynamics across different $H$, we identify a hierarchical structure governing SOT-driven behavior and use it to predict the dynamics of hopfions with general $H$. Furthermore, we show that by appropriately scheduling the time dependence of the SOT, it is possible to repeatedly induce both splitting and recombination of hopfions. These results demonstrate the controllability of hopfion topology via SOT and suggest a pathway toward multilevel spintronic devices based on topology switching.
\end{abstract}
\maketitle

\section{Introduction}
\label{sec:introduction}
The realm of topological magnetism has rapidly advanced, aiming to develop cutting-edge functionalities that surpass conventional electronics by harnessing the high controllability and emergent responses rooted in their topological properties. Among various topological magnetic structures, considerable efforts have been devoted to two-dimensional (2D) magnetic skyrmions, vortexlike spin structures characterized by a topological invariant known as the skyrmion number $N_{\rm sk}$. The skyrmion was originally proposed in high-energy physics as a solitonic solution describing baryons~\cite{Skyrme1961,Skyrme1962}, and its concept was subsequently applied to a variety of physical systems~\cite{Bogdanov1989,Roßler2006,Brey1995,Ho1998,Fukuda2011,Tsesses2018}. Skyrmions in chiral magnets such as B20 compounds~\cite{Muhlbauer2009,Yu2010,Yu2011}, stabilized by the competition between ferromagnetic exchange and Dzyaloshinskii-Moriya interactions~\cite{Dzyaloshinsky1958,Moriya1960}, have been established as a platform for exploring novel emergent phenomena, including their dynamics and transport. Of particular interest is the ultra-low current controllability~\cite{Jonietz2010,Yu2012} and the topological responses arising from emergent magnetic fields generated by noncoplanar spin configurations~\cite{Neubauer2009}. These features have been the focus of extensive research, highlighting the potential of skyrmions in future spintronic applications~\cite{Nagaosa2013}.

With the overall landscape of topological magnetism in 2D systems becoming increasingly well understood, the exploration of novel emergent phenomena and dynamics unique to three-dimensional (3D) topological magnetic structures has emerged as a compelling next challenge. At this pivotal stage in the study of topological magnetism, 3D magnetic hopfions—topological solitons intimately related to knot theory in mathematics—are attracting growing interest. Hopfion counterparts originating from this new class of topology have been found in high-energy physics~\cite{Faddeev1997}, optics~\cite{Sugic2021,Shen2023}, superfluid ${}^3$He~\cite{Volovik1977}, triplet superconductors~\cite{Babaev2002}, bosonic systems~\cite{Kawaguchi2008,Hall2016}, ferroelectrics~\cite{Lukyanchuk2020}, and liquid crystals~\cite{Ackerman2017_nmat,Ackerman2017_PRX,Tai2019}. Lately, magnetic hopfions have also garnered increasing attention owing to their long-awaited experimental observations~\cite{Kent2021,Yu2023,Zheng2023}.

\begin{figure}[t!]
  \centering
  \includegraphics[width=\hsize]{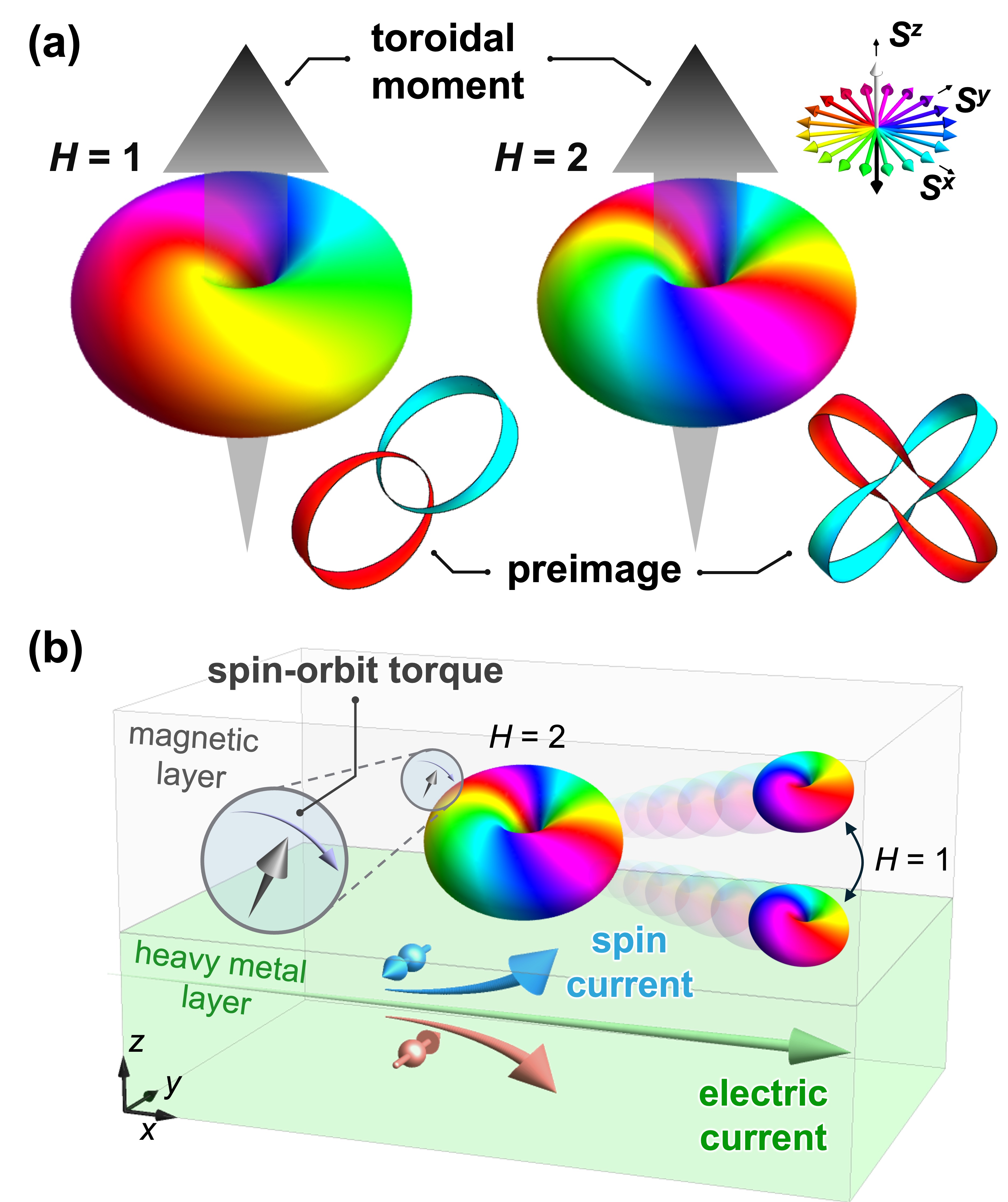}
  \caption{Schematics of the magnetic hopfions and the setup for this study. (a)~Magnetic hopfion with Hopf number $H = 1$ (left) and $H=2$ (right). The gray arrows represent the toroidal moments [Eq.~\eqref{eq:toroidal}]. The lower right insets in left and right panels display their two preimages linked with each other once and twice, respectively. The colors represent spins, with their $S^x$ and $S^y$ $(S^z)$ components indicated by color (grayscale), as shown in the upper right inset. We use this color scheme throughout this paper. (b)~A typical setup for this study. An electric current in the lower heavy metal layer with strong spin-orbit coupling (green arrow) generates a spin current in a perpendicular direction (blue and orange arrows), injected into the upper magnetic layer. The spin current exerts the SOT on the magnetic moments in the magnetic layer. The schematic shows a splitting of the hopfion with $H=2$ into two individual hopfions with $H=1$ by the SOT.}
  \label{schematic}
\end{figure}

A magnetic hopfion can be regarded as a 3D structure composed of a closed loop of a twisted magnetic skyrmion string. Its shape resembles a torus, as exemplified in Fig.~\ref{schematic}(a). Cross sections along the central axis of the torus and perpendicular to it accommodate a skyrmion-antiskyrmion pair and a 2D skyrmionium (a skyrmion nested by an antiskyrmion), respectively. The topological nature of the hopfion is characterized by a topological invariant known as the Hopf number $H$, which is deeply rooted in knot theory~\cite{Hopf1931,Whitehead1947}. Unlike skyrmions, hopfions do not generate a net emergent magnetic field; however, the twisted skyrmion string generates a circulating emergent magnetic field, which in turn induces a toroidal moment oriented perpendicular to the hopfion torus. This distinctive feature gives rise to unique transport properties of conduction electrons and magnons, which are not realized in other magnetic structures~\cite{Pershoguba2021,Liu2022,Saji2023}.

The stabilization mechanisms of hopfions have been investigated theoretically, providing insights into the conditions under which these topological structures can persist. Given that all experimentally observed hopfions to date have been found in slab-shaped samples, surface perpendicular magnetic anisotropy in confined geometries is considered to play an important role in their stabilization~\cite{Sutcliffe2018,Liu2018,Tai2018,Li2022}. In addition, to explore the stability in bulk systems, spin models with competing interactions have also been investigated~\cite{Bogolubsky1988,Sutcliffe2017_frustration,Rybakov2022}. An intriguing feature of these models is the potential stability of hopfions with high Hopf numbers $H \geq 2$. In such models, a hopfion with the Hopf number $H$ follows a sublinear energy bound, $E \geq \eta |H|^{3/4}$, where $\eta$ denotes a model-dependent parameter~\cite{Vakulenko1979}. This relation indicates that multiple hopfions tend to spontaneously merge into a single hopfion whose Hopf number equals the sum of the individual Hopf numbers, consistent with fusion behavior reported in previous studies~\cite{Ward2000, Hietarinta2012, Kasai2025}. In contrast, such spontaneous fusion does not occur in 2D skyrmions; even in analogous 2D frustrate spin models, skyrmions instead tend to stabilize in moleculelike bound structures, such as biskyrmions~\cite{Xichao2017}. This highlights that the rich topological degrees of freedom are a unique feature of 3D hopfions.

Nonequilibrium dynamics of magnetic hopfions under external stimuli has also been the subject of active investigation. For example, in a slab-shaped system, dynamics driven by spin-transfer torque and spin-orbit torque (SOT) under an electric current was theoretically studied, revealing helicity-dependent Hall motion of hopfions, where the helicity of a hopfion is defined as that of the skyrmionium appearing on its cross section~\cite{Wang2019}. In a bulk system, spin-transfer torque was also shown to modulate the shape and orientation of hopfions~\cite{Liu2020}. Furthermore, these investigations were extended to nonlinear dynamics governed by emergent magnetomultipoles~\cite{Liu2022}. However, previous studies have predominantly focused on hopfions with the smallest nontrivial Hopf numbers, $H = \pm1$. Moreover, dynamical interconversion between different $H$ hopfions remains largely unexplored, while recent studies discussed a topological transition between an $H=1$ hopfion and an $H=0$ toron~\cite{Gao2024,Souza2025}. Given the rich topological degrees of freedom associated with hopfions highlighted above, dynamics of $H \geq 2$ hopfions represents a highly compelling research direction.

In this paper, we investigate nonequilibrium dynamics of magnetic hopfions with a wide range of the Hopf number $H$, by numerically solving the Landau-Lifshitz-Gilbert (LLG) equation. In particular, we focus on the effect of SOT originating from a spin current, motivated by the intriguing dynamics of biskyrmions induced by the SOT~\cite{Xichao2017}. First, analyzing the dynamics of hopfions with $H = 1$ over a wide range of the magnetic field and the SOT strength, we demonstrate that the hopfion exhibits translational and precessional motions, which are dependent on their initial orientation and helicity. Next, we perform similar analyses for the hopfions with $H \geq 2$, and find that the SOT acts as an “effective tension” on the hopfions, inducing forced splitting into multiple lower-$H$ hopfions in specific regions of the magnetic field and the SOT strength. We also clarify a hierarchical relation in the splitting dynamics across different Hopf numbers and summarize it in the steady-state phase diagrams for different $H$. Based on these results, we can qualitatively predict the SOT-induced dynamics of high-$H$ hopfions without performing additional simulations. Finally, we demonstrate that the SOT with a designed time dependence can repeatedly cause the splitting and recombination of an $H = 2$ hopfion. Our findings highlight the high tunability of the hopfion’s knot topology and open new avenues for novel memory devices that fully leverage the switching performance.

The rest of the paper is organized as follows. In Sec.~\ref{sec:modelmethod}, we introduce a model Hamiltonian and numerical methods employed in this study. In Sec.~\ref{sec:H1}, we examine the dynamics of the $H = 1$ hopfion~(Sec.~\ref{sec:helicity}) and clarify a steady-state phase diagram summarizing the stability of hopfions in a wide range of the magnetic field and the SOT strength~(Sec.~\ref{sec:H1diagram}). In Sec.~\ref{sec:H2}, we study the SOT-driven dynamics of $H = 2$ hopfions. Notably, we discover an intriguing nonequilibrium dynamics with the intermediate SOT, where the hopfion splits into two hopfions with $H=1$ each~(Sec.~\ref{sec:H2_separation}), and identify the parameter region of such splitting in the steady-state phase diagram~(Sec.~\ref{sec:H2_diagram}). We also discuss the splitting process based on the effective tension picture~(Sec.~\ref{sec:effective_tension}) and provide quantitative estimates of the magnetic field and electric current that are required to cause the separation~(Sec.~\ref{Quantitative estimate}). In Sec.~\ref{sec:H34}, we clarify the SOT-driven dynamics of hopfions with $H = 3$ and $4$ and show that the splitting dynamics of the hopfions also occur for even these higher Hopf numbers. In Sec.~\ref{sec:hierarchy}, we find the hierarchical relations of the hopfion dynamics across the different Hopf numbers and discuss the mechanism underlying the identified dynamics. In Sec.~\ref{sec:recombination}, we demonstrate the splitting and recombination of an $H = 2$ hopfion driven by a time-dependent SOT. Section~\ref{sec:summary} is devoted to the summary of this study.

\section{Model and Method}
\label{sec:modelmethod}
In this section, we introduce the model and methods that we employ in this paper. In Sec.~\ref{sec:model}, we present a spin model with competing interactions. In Sec.~\ref{sec:hopfnumber}, we introduce the Hopf number, which characterizes the topological properties of hopfions, and the toroidal moment, which describes their orientations. In Sec.~\ref{sec:ansatz}, we introduce an ansatz for the spin configuration containing hopfion. In Sec.~\ref{sec:optimization}, we describe the method to prepare the initial states through the gradient-based energy optimization for the hopfions prepared by the ansatz. In Sec.~\ref{sec:LLG}, we introduce the LLG equation used to study the real-time dynamics under the SOT.

\subsection{Model Hamiltonian}
\label{sec:model}
In this study, we consider a spin model with competing interactions on a simple cubic lattice. The Hamiltonian is given by
\begin{align}
\mathcal{H} = -\sum_{\alpha = 1}^4 \sum_{\langle i,j \rangle_\alpha} J_{\alpha}~\bold{S}_i \cdot \bold{S}_j - B \sum_{i} S_i^z,
\label{eq:model}
\end{align}
where $\bold{S}_i = (S_i^x,S_i^y,S_i^z)$ represents the classical spin at site $i$ with $|\bold{S}_i|=1$. The first term represents the exchange interactions up to the fourth-neighbor pairs; the summation with respect to $\langle i,j \rangle_{\alpha}$ runs over $\alpha$th-neighbor pairs. We take $(J_1,J_2,J_3,J_4) = (1,-0.166,0,-0.083)$, and note that similar parameters were employed to stabilize a hopfion~\cite{Bogolubsky1988,Rybakov2022}. The second term describes the Zeeman coupling to an external magnetic field $B$ in units of $g \mu_{\rm B}$, where $g$ is the electron $g$-factor and $\mu_{\rm B}$ is the Bohr magneton. We set the lattice constant to unity and take the system size $N=101^3$, unless otherwise noted, under periodic boundary conditions.

\subsection{Hopf number and toroidal moment}
\label{sec:hopfnumber}
The hopfion is characterized by a topological invariant called the Hopf number $H$, which is defined in continuous space by~\cite{Whitehead1947,Vega1978}
\begin{align}
    H = -\int \bold{B}_{\rm em}(\bold{r}) \cdot \bold{A}_{\rm em}(\bold{r})~d\bold{r} \in \mathbb{Z},
    \label{eq:hopfnum}
\end{align}
where $\bold{B}_{\rm em}(\bold{r})$ is the emergent magnetic field and $\bold{A}_\mathrm{em}(\bold{r})$ represents the vector potential satisfying $\nabla\times\bold{A}_\mathrm{em}(\bold{r}) = \bold{B}_\mathrm{em}(\bold{r})$ at position $\bold{r} = (x, y, z)$. The $\alpha$ component of $\bold{B}_\mathrm{em}$ is given by
\begin{align}
    B_{\rm em}^\alpha(\bold{r}) =\frac{1}{8\pi} \varepsilon_{\alpha\beta\gamma}\bold{S}(\bold{r}) \cdot \left\{\partial_{\beta} \bold{S}(\bold{r}) \times \partial_{\gamma} \bold{S}(\bold{r})
    \right\},
    \label{eq:Bem}
\end{align}
with the Levi-Civita symbol $\varepsilon_{\alpha\beta\gamma}$ and classical spin $\bold{S}(\bold{r})$ in continuous space.

The Hopf number has a geometrical meaning derived from the knot theory in mathematics. As shown in Fig.~\ref{schematic}(a), hopfions have twisted spin structures along their ringlike structures. The topological character of such spin structures can be captured by using the preimage which is defined by connecting spins pointing in the same direction. In magnetic hopfions, each preimage forms a 1D closed ring, as shown in the insets of Fig.~\ref{schematic}(a). The Hopf number is equivalent to how many times any pair of independent preimages link with each other. Indeed, for the hopfion with $H=1$ and $2$ in Fig.~\ref{schematic}(a), the preimages link with each other once and twice, respectively, corresponding to their Hopf numbers.

As we consider the model in Eq.~\eqref{eq:model} on a discrete lattice, it is necessary to replace the integral in Eq.~\eqref{eq:hopfnum} by a discrete summation on the lattice sites. To compute the emergent magnetic field $\bold{B}_{\rm em}$, we replace in Eq.~\eqref{eq:Bem} with the solid angle subtended by a triad of adjacent spins, as given by
\begin{align}
  B_{\rm em}^{\alpha}(\bold{r}_i) = \frac{1}{8\pi}\varepsilon_{\alpha\beta\gamma}[\Omega(\bold{S}_i, &\bold{S}_{i+\beta}, \bold{S}_{i+\gamma}) \nonumber \\
  & + \Omega(\bold{S}_{i+\gamma}, \bold{S}_{i+\beta}, \bold{S}_{i+\beta+\gamma})],
  \label{eq:solidangle}
\end{align}
where $\bold{r}_i$ denotes the position of spin $\bold{S}_i$, and $i+\beta$ denotes the neighboring site to site $i$ in the $\beta$ direction; $\Omega$ is the solid angle calculated by~\cite{Berg1981}
\begin{align}
  &\Omega(\bold{S}_i,\bold{S}_j,\bold{S}_k) \nonumber \\
  & \qquad = 
  2\arctan\left\{
  \frac{\bold{S}_i \cdot (\bold{S}_j \times \bold{S}_k)}{1 + \bold{S}_i\cdot\bold{S}_j + \bold{S}_j\cdot\bold{S}_k + \bold{S}_k\cdot\bold{S}_i}
  \right\},
\end{align}
where the arctangent on the right-hand side is implemented by the atan2 function to satisfy $\Omega \in [-2\pi, 2\pi)$. We calculate the discretized vector potential $\bold{A}_{\rm em}(\bold{r}_i)$ as the inverse Fourier transform of~\cite{Moore2008,Liu2018}
\begin{align}
  \bold{A}_{\rm em}(\bold{k}) = -i \frac{\bold{k} \times \bold{B}_{\rm em}(\bold{k})}{2\pi \bold{k}^2},
\end{align}
where $\bold{k}$ is the momentum vector and $\bold{B}_{\rm em}(\bold{k})$ is the Fourier transform of Eq.~\eqref{eq:solidangle}. Using these discretized forms, we compute the Hopf number density $\rho_{\rm H}(\bold{r}_i) = -\bold{B}_{\rm em}(\bold{r}_i) \cdot \bold{A}_{\rm em}(\bold{r}_i)$ and the Hopf number $H = \sum_i \rho_{\rm H}(\bold{r}_i)$ on a lattice system.

Hopfions are 3D structures with orientational degree of freedom. To characterize the orientation, we introduce the toroidal moment as~\cite{Pershoguba2021,Liu2022}
\begin{align}
    \bold{T}_{\rm em} = \frac{1}{2} \sum_i (\bold{r}_i - \bold{r}_{\rm H}) \times \bold{B}_{\rm em}(\bold{r}_i),
\label{eq:toroidal}
\end{align}
where $\bold{r}_{\rm H}$ is the central position of the hopfion obtained by
\begin{align}
    \bold{r}_{\rm H} = \frac{\sum_i \bold{r}_i \rho_{\rm H}(\bold{r}_i)}{\sum_i \rho_{\rm H}(\bold{r}_i)}.
\label{eq:center}
\end{align}
As shown in Fig.~\ref{schematic}(a), the toroidal moment points in the direction perpendicular to the hopfion's torus, thereby characterizing its orientation. The toroidal moment is useful for the hopfions with $H=1$ and $2$ because they have highly symmetric tori in their optimal forms~\cite{Rybakov2022}, but it is elusive whether it would work for other less symmetric hopfions with higher Hopf numbers, particularly during their dynamics. In fact, as will be noted later, the toroidal moment is not useful for $H = 4$ even in a static environment.

\subsection{Hopfion ansatz}
\label{sec:ansatz}
To prepare an initial spin configuration containing the magnetic hopfion, we adopt the following ansatz in continuous space, which represents a hopfion with Hopf number $H=1$ centered at the origin in a ferromagnetic background~\cite{Wang2019}:
\begin{align}
  \begin{split}
      S^x (\bold{r}) &= \frac{-4[2xz + y(r^2-1)]}{(1 + r^2)^2},\\
      S^y (\bold{r}) &= \frac{-4[2yz - x(r^2-1)]}{(1 + r^2)^2},\\
      S^z (\bold{r}) &= 1 - \frac{8(x^2 + y^2)}{(1 + r^2)^2},
  \label{eq:ansatz}
  \end{split}
\end{align}
where $r^2 = |\bold{r}|^2$, and extract the values at each lattice site. The above ansatz satisfies $\bold{S} \to (0,0,1)$ as $r \to \infty$, representing the ferromagnetic background. A hopfion with an arbitrary number of Hopf number $H$ can be obtained by the transformation as~\cite{Hietarinta1999}
\begin{align}
  r e^{i \theta} \to r e^{i H \theta},~ z \to z,
  \label{eq:arbHopfnum}
\end{align}
where $(r, \theta)$ is the polar coordinate of $(x,y)$. The isosurfaces of $S^z=0$ for the hopfions with $H = 1$ and $2$ are displayed in Fig.~\ref{schematic}(a). Note that the ideal isosurface depicted from Eqs.~\eqref{eq:ansatz} and \eqref{eq:arbHopfnum} has a toroidal shape, but it is modified by the energy optimization in the next subsection.

\subsection{Energy optimization}
\label{sec:optimization}
We prepare the initial state for the real-time dynamics simulation, optimizing the spin configuration in the ansatz by the gradient descent method using the Adam~\cite{Kingma2014}. To optimize not only the hopfion's shape but also its orientation under the Hamiltonian in Eq.~\eqref{eq:model}, we perform the energy optimization for hopfions prepared with random orientations. Such states can be prepared by applying rotation matrices, parameterized by the Euler angles $\alpha, \beta,$ and $\gamma$, to $\bold{r}$ in Eq.~\eqref{eq:ansatz}; we calculate $\tilde{\bold{r}} = \hat{R}_z(\gamma) \hat{R}_x(\beta) \hat{R}_z(\alpha)\bold{r}$ where $\hat{R}_\mu(\theta)$ is the rotation matrix around a $\mu$ axis by $\theta$, and obtain $\bold{S}(\tilde{\bold{r}})$ from Eq.~\eqref{eq:ansatz} with the rotated toroidal moment in Eq.~\eqref{eq:toroidal}. After the transformation, we discretize the ansatz as $\bold{S}(\bold{r}) \to \bold{S}(\bold{r}_i)$ to place a hopfion on a lattice. We prepare 30 sets of random Euler angles for each case, and perform energy optimization starting from the rotated hopfions.

For the hopfions with $H = 1$, we find that the toroidal moments of the optimized hopfions are directed to three independent easy axes parallel to the $x, y,$ and $z$ axes. In contrast, those for $H = 2$ and $3$ point to four independent easy axes along the $[111]$ directions. We adopt hopfions with toroidal moments of $\bold{T}_{\rm em} \parallel (0,0,1)$ for $H = 1$ and $\bold{T}_{\rm em} \parallel (1,1,1)$ for both $H = 2$ and $3$ as the initial states for LLG simulations in the following sections. For each case, we select the set of Euler angles from the 30 trials that converges most quickly to the corresponding easy axis, and perform $5 \times 10^5$ steps of gradient descent optimization to obtain the initial states for the LLG simulations. For hopfions with $H=4$, we were not able to find preferable easy axes for the toroidal moments after the energy optimization. This is perhaps because the hopfion with $H=4$ has a complicated torus shape~\cite{Rybakov2022}. Therefore, we assume an initial state with $\bold{T}_{\rm em} \parallel (0,0,1)$ in Sec.~\ref{sec:H4_separation}. In Sec.~\ref{sec:hierarchy}, we also adopt other initial states with different $\bold{T}_{\rm em}$ to carefully investigate the SOT dependence of the dynamics.

\subsection{Landau-Lifshitz-Gilbert equation}
\label{sec:LLG}
To elucidate the real-time dynamics of hopfions, we employ the LLG equation given in a dimensionless form as
\begin{align}
  \frac{d \bold{S}_i}{d\tau} = \bold{S}_i \times \bold{H}_i^{\rm eff} + \alpha \bold{S}_i \times \frac{d \bold{S}_i}{d\tau} + \zeta \bold{S}_i \times (\bold{S}_i \times \bold{p}),
  \label{eq:LLG}
\end{align}
where $\tau$ denotes dimensionless time, and $\alpha$, $\zeta$, and $\bold{p}$ represent the Gilbert damping, the strength of the dampinglike SOT, and the polarization direction of the spin current generating the SOT, respectively. The effective field $\bold{H}_i^{\rm eff}$ in Eq.~\eqref{eq:LLG} is defined as $\bold{H}_i^{\rm eff} = \frac{\partial \mathcal{H}}{\partial \bold{S}_i}$. We numerically solve Eq.~\eqref{eq:LLG} by using the fourth-order Runge-Kutta method. We switch on the SOT at time $\tau = 0$, unless otherwise noted. We take the Gilbert damping $\alpha = 0.2$, the time step $\Delta \tau = 0.1$, and $\bold{p} = (0,1,0)$.

\section{Results}
\label{sec:result}
In this section, we investigate the dynamics of hopfions with various Hopf numbers $H$ driven by the SOT, solving the LLG equation given in Eq.~\eqref{eq:LLG}. Section~\ref{sec:H1} corresponds to $H=1$, Sec.~\ref{sec:H2} to $H=2$, and Sec.~\ref{sec:H34} to $H=3$ and $H=4$. In Sec.~\ref{sec:hierarchy}, we compare these results and discuss a hierarchical structure in the SOT-driven dynamics of hopfions with different $H$.

\subsection{SOT-driven dynamics of $H=1$ hopfion}
\label{sec:H1}
First, we elucidate the nonequilibrium dynamics of a hopfion with $H = 1$ driven by the SOT. In Sec.~\ref{sec:helicity}, we show that the hopfion exhibits translational and precessional motions depending on its helicity. In Sec.~\ref{sec:H1diagram}, we present the steady-state phase diagram while changing the strengths of the magnetic field $B$ and the SOT $\zeta$.

\subsubsection{Translational and precessional motion depending on helicity}
\label{sec:helicity}
\begin{figure*}[tb]
  \centering
  \includegraphics[width=\hsize]{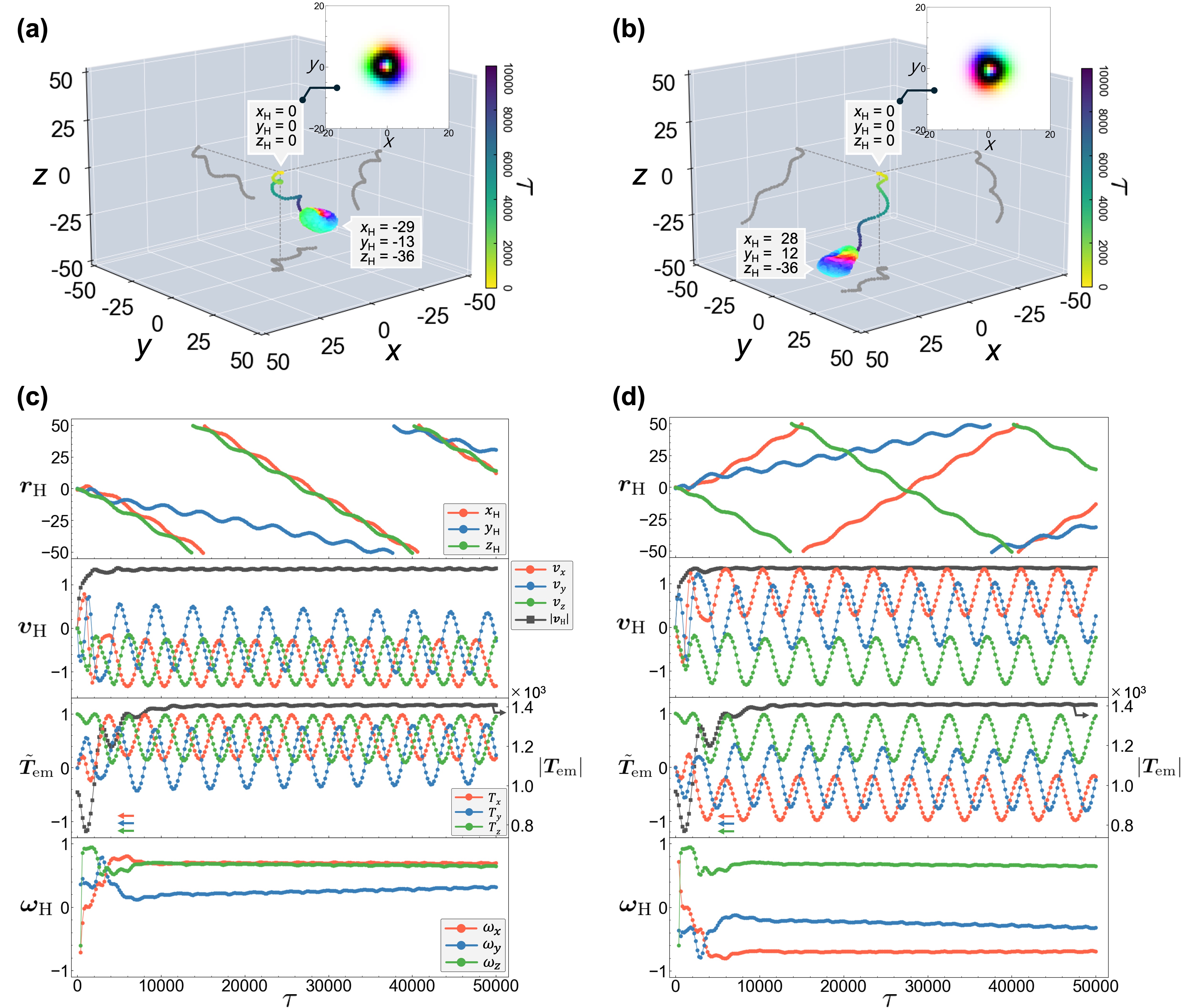}
  \caption{SOT-driven dynamics of the hopfion with $H = 1$ under $\zeta = 0.001$ and $B = 0.003$, starting from the initial state with (a) the helicity $\eta=0$ and (b) $\eta=\pi$. The insets show the 2D spin structures on the slice at $z=0$ of the initial state. The colored and gray curves denote the real-space trajectory and its projections onto each boundary plane up to $\tau = 10000$, respectively. In (a) and (b), the hopfion reaches (-29, -13, -36) and (28, 12, -36) at $\tau=10000$, respectively. The lower panels display time evolutions of the hopfion's position, velocity, toroidal moment, and precessional axis of the toroidal oscillation during the dynamics up to $\tau=50000$.}
  \label{fig2:H1_dynamics}
\end{figure*}

Figure~\ref{fig2:H1_dynamics}(a) shows the trajectory of the hopfion with $H=1$ in real space, starting from the initial state prepared by the methods in Secs.~\ref{sec:ansatz} and \ref{sec:optimization}. Here, we take $\zeta=0.001$ and $B=0.003$, and initialize the position of the hopfion, $\bold{r}_{\rm H} = (x_{\rm H}, y_{\rm H}, z_{\rm H})$, at the origin. We find that the hopfion is moved by the SOT in a specific direction with meandering and reaches $\bold{r}_{\rm H} = (-29, -13, -36)$ at $\tau=10000$. The upper two panels of Fig.~\ref{fig2:H1_dynamics}(c) display the time evolutions of $\bold{r}_{\rm H}$ and the velocity $\bold{v}_{\rm H} = (v_x,v_y,v_z)$ during the dynamics. Here, the velocity is measured by
\begin{align}
    \bold{v}_{\rm H}(\tau) = \bold{r}_{\rm H}(\tau) - \bold{r}_{\rm H}(\tau - \Delta \tau),
\label{eq:velocity}
\end{align}
with $\Delta \tau = 200$ and $\bold{v}_{\rm H}(0) = \bold{0}$. The plot of the position shows that the hopfion exhibits a steady translational motion with slight oscillations in all components. Correspondingly, all components of the velocity also oscillate with weak time dependence on their amplitudes after the initial acceleration by the SOT. The third panel of Fig.~\ref{fig2:H1_dynamics}(c) displays the time evolution of the normalized toroidal moment, $\tilde{\bold{T}}_{\rm em} = \bold{T}_{\rm em}/|\bold{T}_{\rm  em}| = (T_x,T_y,T_z)$. The toroidal moment also exhibits oscillations in all components, indicating a precession of a hopfion orientation. We calculate the precessional axial vector from the set of normalized toroidal moments at three consecutive times, which is given by
\begin{align}
  \bold{\omega}_{\rm H}(\tau)  = \frac{\delta \tilde{\bold{T}}_{\rm em}(\tau - \Delta \tau) \times \delta \tilde{\bold{T}}_{\rm em}(\tau)}{|\delta \tilde{\bold{T}}_{\rm em}(\tau - \Delta \tau) \times \delta \tilde{\bold{T}}_{\rm em}(\tau)|},
\label{eq:precession}
\end{align}
where $\delta \tilde{\bold{T}}_{\rm em}(\tau) = \tilde{\bold{T}}_{\rm em}(\tau) - \tilde{\bold{T}}_{\rm em}(\tau - \Delta \tau)$ with $\Delta \tau = 200$, and plot the time evolution in the lowest panel of Fig.~\ref{fig2:H1_dynamics}(c). In the steady state at $\tau = 50000$, $\bold{\omega}_{\rm H}$ reaches $(0.696, 0.318, 0.644)$, where the $x$ and $y$ components have nearly converged, but the $z$ component still shows a slight time dependence. Thus, the hopfion exhibits not only translational but also precessional motion. From a Fourier analysis for $\tilde{\bold{T}}_{\rm em}$, we obtain the precession period as $\simeq 4560$. We note that, during the early stage of the dynamics, the SOT induces deformation of the hopfion and it causes the increasing norm of the toroidal moment, but after a long time, the norm saturates at a constant with slight oscillations. 

The motion of the $H=1$ hopfion depends on its helicity. This is shown in Figs.~\ref{fig2:H1_dynamics}(b) and \ref{fig2:H1_dynamics}(d). Here, we prepare the initial state by rotating that used in Fig.~\ref{fig2:H1_dynamics}(a) by $\pi$ around the $z$ axis, resulting in the $\pi$ difference in the helicity; see the 2D slices at $z=0$ in the insets of the top panels of Figs.~\ref{fig2:H1_dynamics}(a) and \ref{fig2:H1_dynamics}(b), where we define the helicity as $\eta=0$ and $\pi$, respectively. Compared to the results of Fig.~\ref{fig2:H1_dynamics}(a), the hopfion moves in a different direction and reaches $\bold{r}_{\rm H}=(28,12,-36)$ at $\tau=10000$, where the $x$ and $y$ components have reversed signs relative to those in Fig.~\ref{fig2:H1_dynamics}(a). While the direction of the velocity differs from that in Fig.~\ref{fig2:H1_dynamics}(a), its norm is almost the same. The toroidal moment also develops to almost the same value and exhibits a precession with the same oscillation period, while each component show different behavior from that in Fig.~\ref{fig2:H1_dynamics}(a). The precessional axial vector $\bold{\omega}_{\rm H}$ reaches $(-0.695, -0.318, 0.645)$ in the steady state, which is close to that in Fig.~\ref{fig2:H1_dynamics}(a) with the sign reversals of its $x$ and $y$ components. These results indicate that the hopfion with $\pi$ difference in the helicity also shows both translational and precessional motions with the same velocity and oscillation period but different moving direction and precessional axis.

These helicity-dependent motions driven by the SOT are understood from the hopfion's spin structures as follows. As the helicity for the hopfion is defined by that of the skyrmionium found on the $xy$ cut, a rotation of the hopfion around the $z$ axis corresponds to a change in the helicity of the skyrmionium. It is known that a direction of a translational motion of a SOT-driven skyrmionium depends on its helicity~\cite{Xia2020}. This suggests that the direction of the hopfion’s translational motion is determined by the initial helicity of the skyrmionium within the hopfion. This picture is consistent with the fact that only the $x$ and $y$ components of the velocity change their signs in response to a $\pi$-rotation in the helicity, as found in Fig.~\ref{fig2:H1_dynamics}(b). Meanwhile, a perpendicular cut parallel to the toroidal moment includes both skyrmion and antiskyrmion. They show not only a rotation centered at their core but also a global rotation under the SOT~\cite{Shizeng2016, Xichao2017}. Directions of the global rotations are opposite between skyrmion and antiskyrmion. This suggests that the precession of the toroidal moments originates from the competing global rotations of the skyrmion and the antiskyrmions included in the hopfion, potentially explaining the results in Fig.~\ref{fig2:H1_dynamics}(b).

\subsubsection{Steady-state phase diagram}
\label{sec:H1diagram}
\begin{figure}[t!]
  \centering
  \includegraphics[width=\hsize]{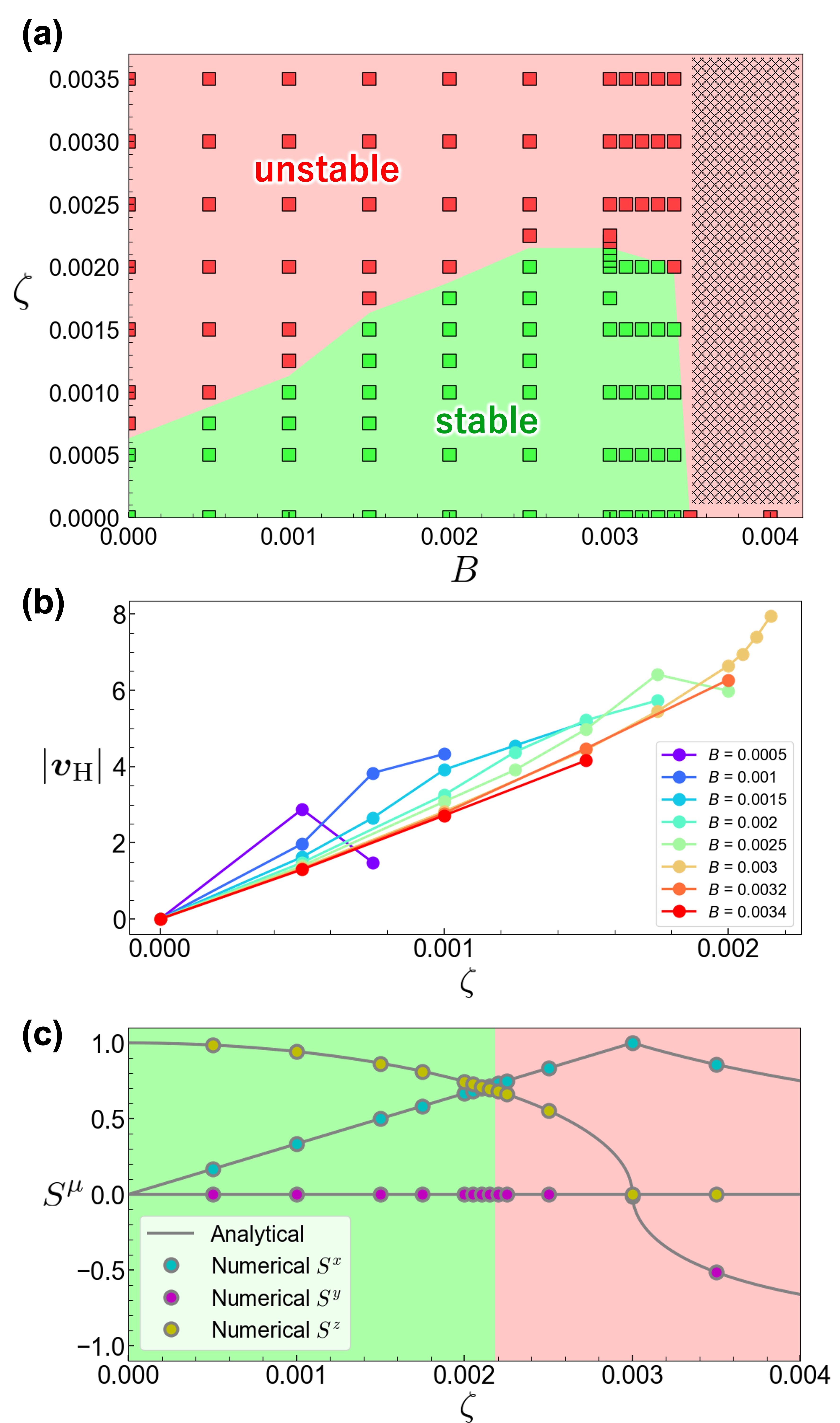}
  \caption{(a)~Steady-state phase diagram of the hopfion with $H = 1$ under the magnetic field $B$ and the SOT with strength $\zeta$. The points at $\zeta = 0$ correspond to the initial states obtained by following the procedures in Secs.~\ref{sec:ansatz} and \ref{sec:optimization}, while the other points correspond to the steady states of the LLG analyses performed using those initial states up to $\tau=50000$. The hatched region for $B \geq 0.0035$ is not accessible due to the instability of the initial state in the absence of the SOT. (b)~Terminal velocity $v$ of the hopfions as a function of $\zeta$ in the stable region, given by the norm of Eq.~\eqref{eq:velocity} with $\tau=50000$ and $\Delta \tau = 1000$. (c)~Averaged magnetization as a function of $\zeta$ at $B=0.003$ and $\tau = 50000$. The colored symbols and the gray lines correspond to the numerical results and the trivial solution in Eq.~\eqref{eq:forcedFM}, respectively. The background colors represent the stable and unstable regions estimated from (a).}
  \label{fig3:H1_diagram}
\end{figure}

We perform LLG analyses in a wide range of the magnetic field $B$ and SOT strength $\zeta$. We find that the $H=1$ hopfion remains topologically stable for small $B$ and $\zeta$, whereas it becomes unstable when $B$ and $\zeta$ become large. Figure~\ref{fig3:H1_diagram}(a) summarizes such behaviors in the steady state after sufficiently long time $\tau = 50000$. In the green shaded region, the hopfion remains topologically stable and exhibits translational and precessional motion, similar to the results shown in Fig.~\ref{fig2:H1_dynamics}. In contrast, in the red shaded region, the hopfion becomes unstable and relaxes to a ferromagnetic state. The hatched area for $B \geq 0.0035$ is not accessible within the present simulations because the initial states are unstable even at $\zeta = 0$.

Figure~\ref{fig3:H1_diagram}(b) shows the norm of the hopfion's velocity, $|\bold{v}_{\rm H}|$, at $\tau = 50000$ as a function of $\zeta$ in the stable region for several selected values of $B$, where the norm is obtained by Eq.~\eqref{eq:velocity} with $\Delta \tau = 1000$. It reveals a tendency that the translational motion becomes faster with increasing $\zeta$, up to the boundary to the unstable region. This suggests that a stronger SOT generates larger driving force for the hopfion, whereas eventually makes it unstable. We also observe that the velocity tends to be slower or even decrease just before $\zeta$ reaches the unstable region. This behavior is likely caused by a severe deformation of the hopfion due to the in-plane tilt of the background magnetization. A more detailed discussion will be given in Sec.~\ref{sec:H2_diagram}, focusing on the results for $H = 2$ hopfions, where this trend is more pronounced.

A trivial steady-state solution can be obtained by assuming a ferromagnetic state in the LLG equation in Eq.~\eqref{eq:LLG}. The solution is given by
\begin{align}
    \bold{S}_i =
    \begin{cases}
        (\zeta/B, 0, \sqrt{1 - (\zeta/B)^2}) &(\zeta \leq B)\\
        (B/\zeta, -\sqrt{1 - (B/\zeta)^2}, 0) &(\zeta > B).
    \label{eq:forcedFM}
    \end{cases}
\end{align}
In this solution, as $\zeta$ increases, the spins tilt towards the $x$ direction, and the system enters a ferromagnetic state with spins aligned within the $xy$ plane when $\zeta$ exceeds $B$, as shown in Fig.~\ref{fig3:H1_diagram}(c) for $B=0.003$. We also plot the numerical values of $S^x$, $S^y$, and $S^z$ averaged over the whole system in the steady state, which agree well with Eq.~\eqref{eq:forcedFM} even in the stable region where the hopfion is present. This indicates that the majority spins in the background of the hopfion follow the trivial solution. It is worth noting that the hopfion instability starts around when $S^x$ becomes comparable to $S^z$, where $\zeta/B = 1/\sqrt2$ from Eq.~\eqref{eq:forcedFM}. This is presumably because the in-plane tiltings of the background spins severely distort the torus shape of the hopfion and finally destroys it when the tilted component becomes comparable to the $z$ component.

\subsection{SOT-driven dynamics of $H=2$ hopfion}
\label{sec:H2}
Next, we consider the dynamics of hopfions with $H=2$. In Sec.~\ref{sec:H2_separation}, we demonstrate that a sufficiently strong SOT induces a splitting of an $H=2$ hopfion into two $H=1$ hopfions. In Sec.~\ref{sec:H2_diagram}, we present the steady-state phase diagram while changing the strengths of the magnetic field $B$ and the SOT $\zeta$. In Sec.~\ref{sec:effective_tension}, we focus on the splitting process and propose an ``effective tension" picture. In Sec.~\ref{Quantitative estimate}, we discuss the conditions under which the splitting occurs in view of the helicity-dependent dynamics of $H=1$ hopfions revealed in Sec.~\ref{sec:helicity}.

\subsubsection{Splitting into two hopfions}
\label{sec:H2_separation}
\begin{figure}[tb]
  \centering
  \includegraphics[width=\hsize]{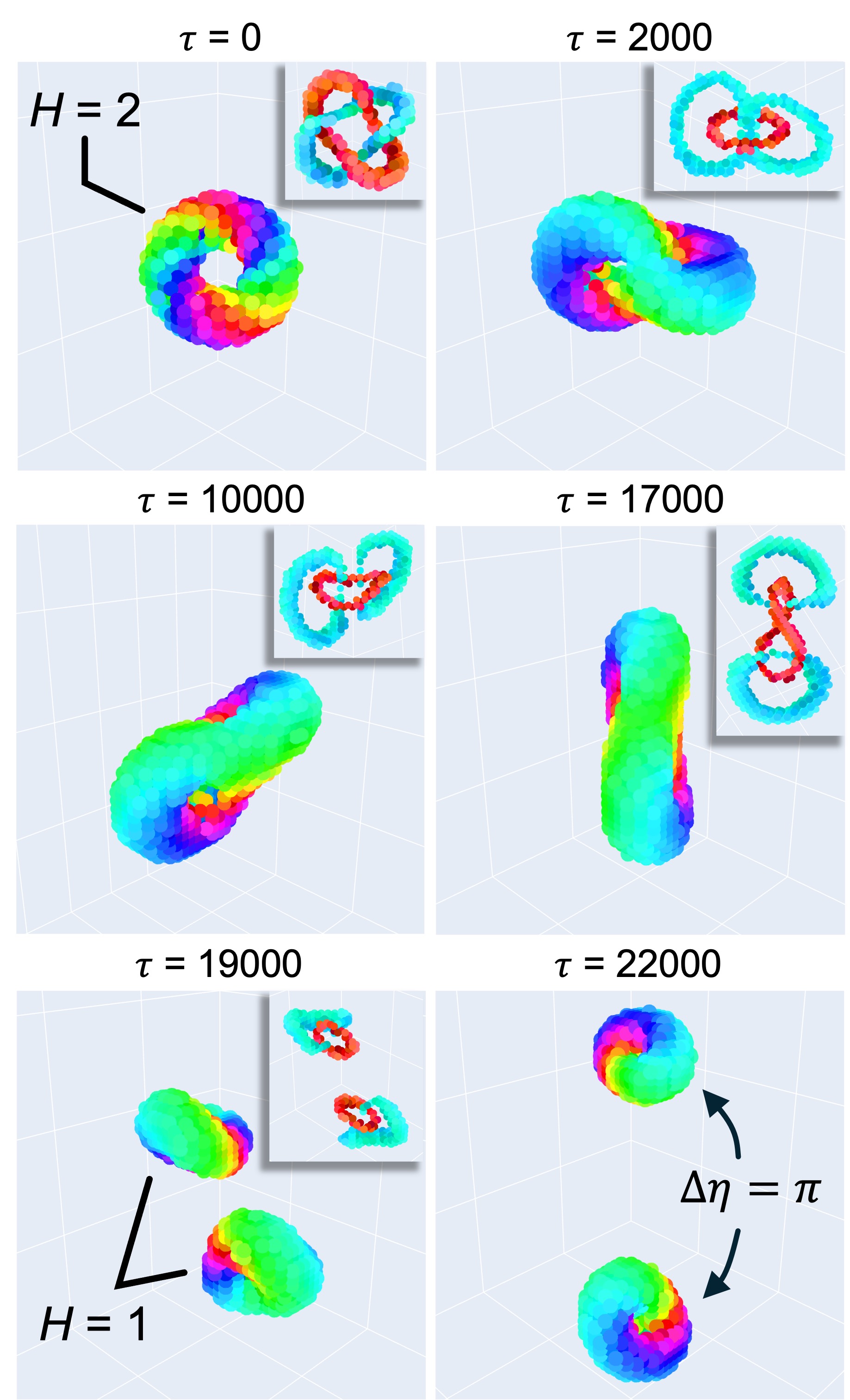}
  \caption{Snapshots of the hopfion with $H=2$ under $B=0.003$ and $\zeta = 0.002$. All panels except the lower-right one include insets showing the preimages on which $S^x \simeq +1$~(blue) or $S^x \simeq -1$~(red). For $\tau \gtrsim 19000$, the hopfion with $H=2$ splits into two hopfions, each with $H=1$, whose helicities $\eta$ differ by $\Delta \eta = \pi$. The viewpoint is varied with time for better visibility.}
  \label{fig4:H2_separation}
\end{figure}

Figure~\ref{fig4:H2_separation} shows snapshots of the $H=2$ hopfion dynamics with $\zeta=0.002$ and $B=0.003$. Starting from the symmetric form at $\tau=0$, the hopfion undergoes a twisting deformation during its time evolution. Interestingly, the blue preimage satisfying $S^x \simeq +1$ develops an ``8"-shape and linked to the red one satisfying $S^x \simeq -1$, as shown in the snapshot at $\tau=2000$. Around $\tau = 10000$, the junction of the ``8"-shaped preimage starts to split, and the blue preimage splits into two isolated rings. In contrast, the red preimage remains a single ring, but it starts to twist into another ``8"-shaped loop, as shown in the snapshot at $\tau=17000$. At this stage, the spin state still retains the hopfion with an elongated but single bound structure. However, by $\tau = 19000$, the hopfion is split into two isolated hopfions with $H = 1$ each. The two hopfions have different helicities by $\pi$, and thus exhibit different translational and precessional motions after the splitting, similar to the dynamics revealed in Sec.~\ref{sec:helicity}.

We note that similar splitting by the SOT was reported for the biskyrmion~\cite{Xichao2017}, a molecularlike structure formed by two skyrmions bound at a finite distance. However, a key difference lies in the fact that despite being a single symmetric topological object, the hopfion with $H=2$ can be split by the SOT.

\subsubsection{Steady-state phase diagram}
\label{sec:H2_diagram}
\begin{figure}[tb]
  \centering
  \includegraphics[width=\hsize]{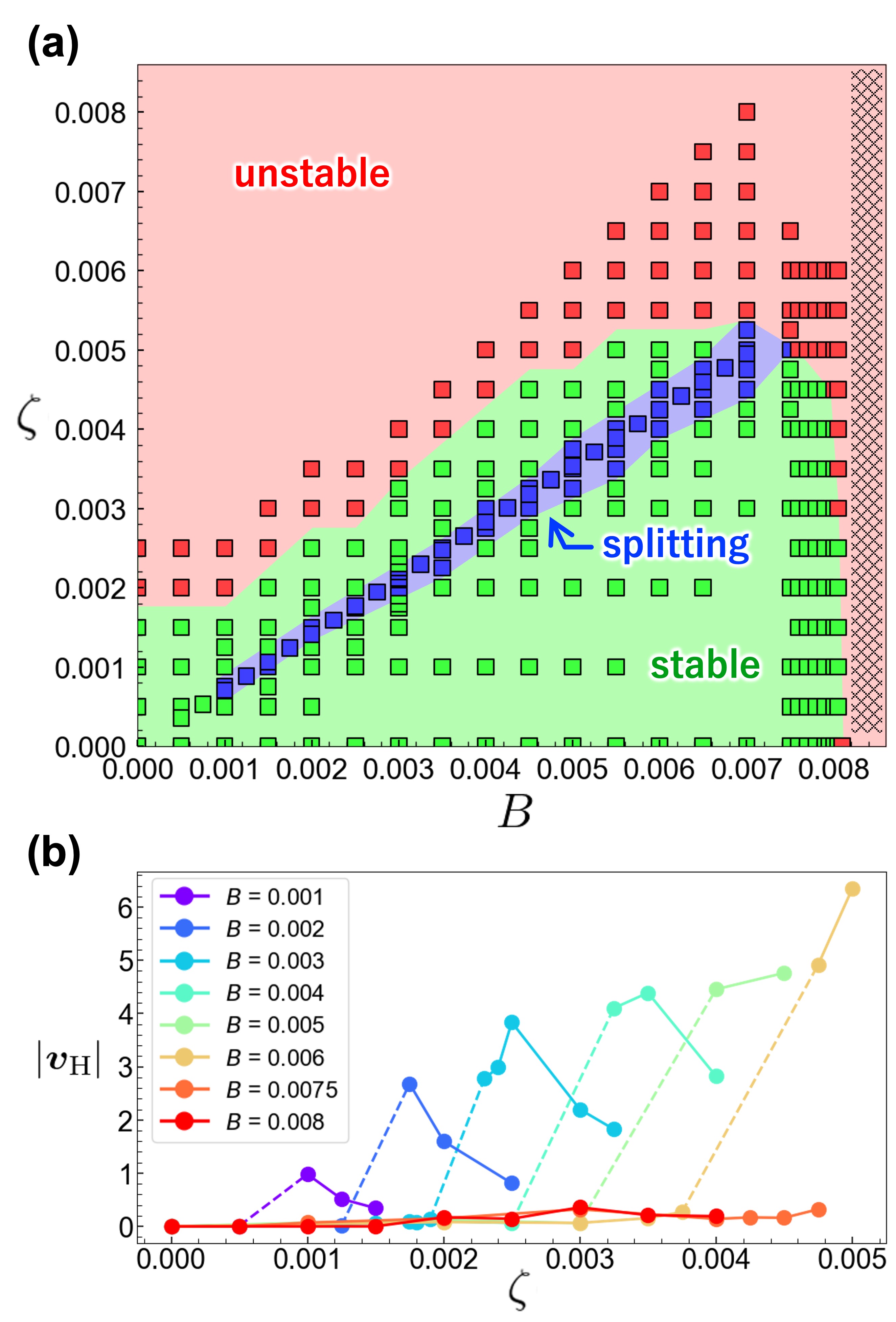}
  \caption{(a)~Steady-state phase diagram of the hopfion with $H = 2$. The notations are common to Fig.~\ref{fig3:H1_diagram}(a). The splitting of the hopfion into two individuals with $H=1$ occurs in the blue shaded region. (b)~Terminal velocity $v$ of the hopfions as a function of $\zeta$ in the stable region, obtained by the same manner as in Fig.~\ref{fig3:H1_diagram}(b). The dashed lines represent the $\zeta$ range where splitting occurs.}.
  \label{fig5:H2_diagram}
\end{figure}

Performing LLG analyses in a wide range of $B$ and $\zeta$ systematically, we summarize the steady-state phase diagram after sufficiently long time $\tau = 50000$ in Fig.~\ref{fig5:H2_diagram}(a). This diagram has four distinct regions. In the red shaded region, the $H = 2$ hopfion becomes unstable and relaxes into a ferromagnetic state, similar to the corresponding region for the $H = 1$ hopfion in Fig.~\ref{fig3:H1_diagram}(a). It is worth noting that the $H = 2$ hopfion can be topologically stable against a stronger $\zeta$ compared to that with $H = 1$. This suggests that the complexity in the linking of the preimages enhances the robustness of the $H = 2$ hopfions. In the green shaded region, the hopfion maintains a single structure with $H=2$, but the region is further split into two by the blue shaded region. 

In the blue shaded region, the $H=2$ hopfion splits into two with $H=1$ each, as shown in Fig.~\ref{fig4:H2_separation}. This region extends around the line $\zeta/B = 1/\sqrt{2}$, where the $x$ and $z$ components of the background magnetization are equal in the trivial solution in Eq.~\eqref{eq:forcedFM}. This suggests that the splitting of the hopfion is triggered by the SOT which causes the in-plane tilt of the background spins up to $S^x \simeq S^z$.

In the green shaded regions below and above the blue one, the $H=2$ hopfion remains topologically stable as an isolated soliton, but shows different dynamics. Let us first discuss the lower green region. In this region, the hopfion shows no translational and precessional motion despite slight deformation, in contrast to the $H=1$ case. The suppression of the translational motion is shown in Fig.~\ref{fig5:H2_diagram}(b) as $|\bold{v}_{\rm H}| \simeq 0$. This suggests that the energy of the SOT is insufficient to drive the hopfion in the lower green region. In contrast, in the upper green region above the blue one, the hopfion has nonzero velocity, as shown in Fig.~\ref{fig5:H2_diagram}(b), with a precessional motion around the toroidal moment (not shown). This indicates that the energy of the SOT is consumed mainly for the translational and precessional motions, rather than the splitting in the adjacent blue region. We note that as shown in Fig.~\ref{fig5:H2_diagram}(b), the velocity tends to decrease as $\zeta$ approaches the unstable region. A similar trend, albeit weaker, was also observed for the $H=1$ case in Fig.~\ref{fig3:H1_diagram}(b). For such $\zeta$, the background magnetization is mostly oriented in-plane as indicated by Eq.~\eqref{eq:forcedFM}, hence the hopfion is significantly distorted, possibly causing the decrease in the velocity; see also the discussion for Fig.~\ref{fig6:H2_tension}(a).

\subsubsection{Effective tension picture}
\label{sec:effective_tension}
\begin{figure*}[tb]
  \centering
  \includegraphics[width=\hsize]{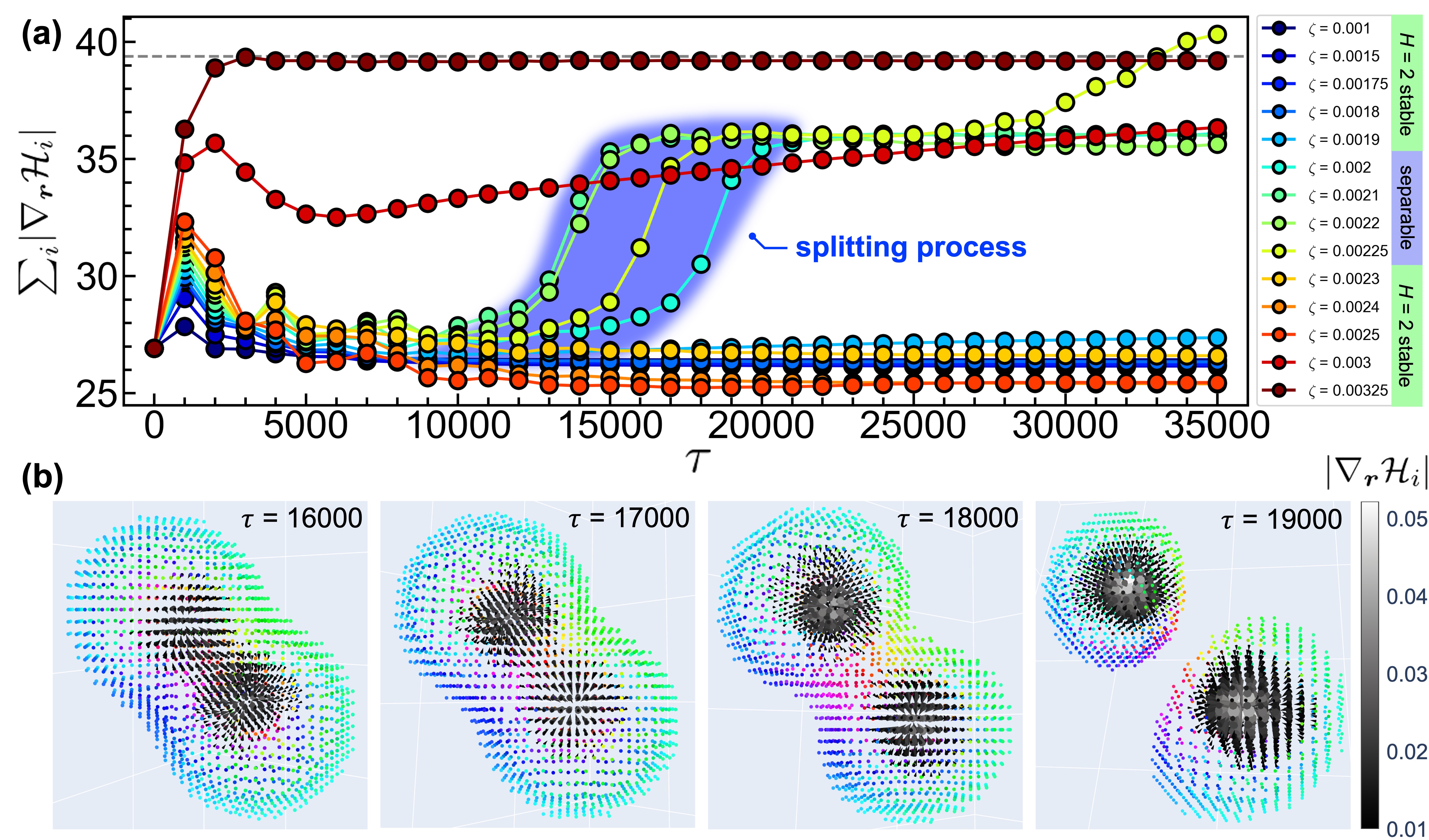}
  \caption{(a)~Time evolution of the absolute sum of the spatial gradient of energy density at $B = 0.003$ with variations of the SOT strength $\zeta$. Splitting occurs in the blue-shaded time regions, during which the sum of the spatial gradient rapidly increases. The horizontal gray dashed line is the reference corresponding to twice the value of the gradient sum of a relaxed hopfion ($H=1$) in the absence of the SOT. (b)~Snapshots of the dynamics around the splitting under $\zeta = 0.002$. The colored dots represent the spins at each site. The arrows represent the gradient vectors $\nabla_{\bold r} \mathcal{H}_i$ and their norms are represented by the grayscale bar in (b).}
  \label{fig6:H2_tension}
\end{figure*}

To clarify the mechanism of splitting, we examine the deformation of the $H=2$ hopfion during the dynamics in detail. We assess the deformation by calculating the spatial gradient of the energy at each site, $\nabla_{\bold r} \mathcal{H}_i$, where $\mathcal{H}_i$ represents the energy at each site $i$ as 
\begin{align}
    \mathcal{H}_i = -\frac{1}{2}\sum_{\alpha = 1}^4 \sum_{j_{\alpha}} J_{\alpha}~\bold{S}_i \cdot \bold{S}_{j_{\alpha}} - B S_i^z,
\label{eq:Hi}
\end{align}
and the spatial derivative $\nabla_{\bold r} = (\partial_x,\partial_y,\partial_z)$ is computed as $\partial_{\alpha} \mathcal{H}_i = (\mathcal{H}_{i+\alpha} - \mathcal{H}_{i-\alpha})/2$ on the discrete lattice. In Eq.~\eqref{eq:Hi}, the summation with respect to $j_{\alpha}$ is taken over the spins at $\alpha$th-neighbor sites of $i$. Figure~\ref{fig6:H2_tension}(a) shows the time evolution of the absolute sum of the gradient, $\sum_i |\nabla_{\bold r} \mathcal{H}_i|$, at $B = 0.003$. In the stable region for small $\zeta$, it is commonly observed that the gradient value converges to around $25 \sim 28$, with weak time dependence. In contrast, in the region where the hopfions split, the dynamics presents the gradient around $27$ in the early stage, but the value rapidly increases in the time range shaded in blue that corresponds to the splitting, and eventually converges to around $36$ after the splitting~\cite{note_upturn}. We note that the value after the splitting is comparable to twice the gradient value of the $H=1$ hopfion in the absence of the SOT, which is indicated by the gray dashed line at $\simeq$ 39 in Fig.~\ref{fig6:H2_tension}(a). These results indicate that the spatial gradient of the energy is a good indicator to detect the change of the linking topology of the hopfions toward splitting.

To more closely examine the splitting process, we show in Fig.~\ref{fig6:H2_tension}(b) snapshots of hopfions before and after the splitting at $\zeta=0.002$. Here, the arrows represent the local energy gradient $\nabla_{\bold r} \mathcal{H}_i$ with the grayscale bar in the right, and the colored dots represent the spins satisfying $S^z \simeq 0$. Just before the splitting shown in the left two panels of Fig.~\ref{fig6:H2_tension}(b), the $H=2$ hopfion is deformed so that one of two preimages is disconnected, as shown in the middle row of Fig.~\ref{fig4:H2_separation}. In these states, the energy gradient is concentrated around the hopfion center, suggesting that an ``effective tension" acts between disconnected preimages. As time evolves, the gradient becomes dilute around the junction, and is concentrated at the center of each $H=1$ hopfion after the splitting, as shown in the right two panels of Fig.~\ref{fig6:H2_tension}(b). This indicates that strong SOT provides sufficient energy to overcome the effective tension binding the hopfions. After the splitting, the energy exerted by the SOT on the hopfions is stored within each hopfion. The increase in the gradient values around the blue-shaded time region in Fig.~\ref{fig6:H2_tension}(a) can be regarded as a ``work function" required to split $H=2$ into two $H=1$.

Once $\zeta$ reaches the upper stable green region above the splitting blue one in Fig.~\ref{fig5:H2_diagram}(a), the gradient value again converges to around $25 \sim 28$, as shown in Fig.~\ref{fig6:H2_tension}(a). However, near the boundary of the unstable region, the hopfions become significantly distorted, leading to an increase in the gradient value, as exemplified at $\zeta=0.003$ (red) and $\zeta=0.00325$ (brown). This severe distortion may account for the reduction in velocity observed in Fig.~\ref{fig5:H2_diagram}(b).

\begin{figure}[t!]
  \centering
  \includegraphics[width=\hsize]{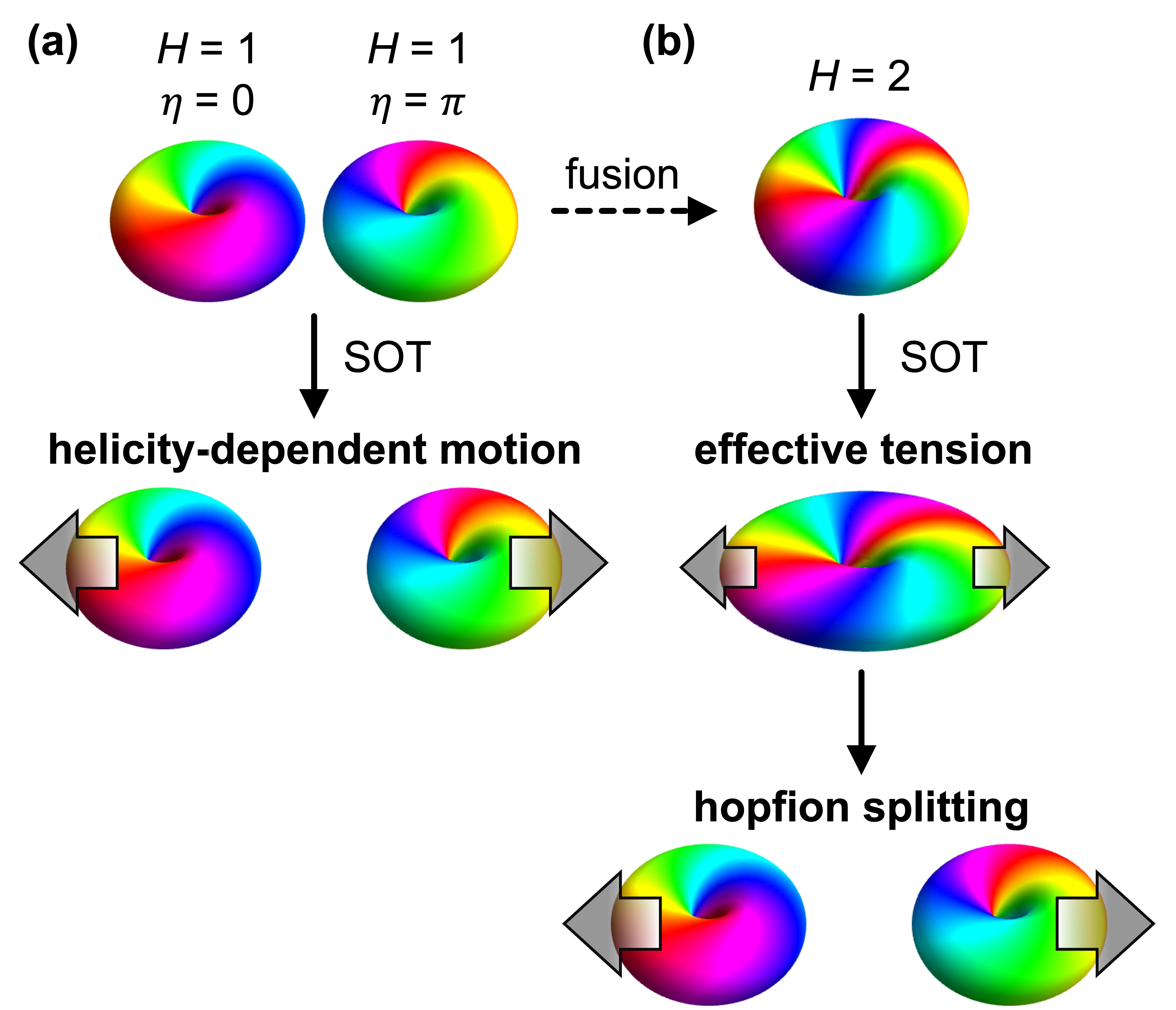}
  \caption{(a)~A schematic illustration of SOT-driven motion of $H=1$ hopfions. Two hopfions with a helicity diffence $\Delta \eta = \pi$ exhibit translational motions in the opposite directions, as demonstrated in Fig.~\ref{fig2:H1_dynamics}. Without the SOT, horizontally placed two $H=1$ hopfions experience an attractive interaction, resulting in their fusion into an $H=2$ hopfion, indicated by the black dashed line. (b)~A schematic illustration of SOT-driven dynamics of the $H=2$ hopfion. Reflecting the helicity-dependent motion in the $H=1$ case, the SOT works as an effective tension, which eventually leads to splitting into two $H = 1$ hopfions.}
  \label{fig_separation}
\end{figure}

Combining the effective tension picture introduced above with the results of $H = 1$ hopfions found in Sec.~\ref{sec:H1}, let us discuss the qualitative mechanism of the SOT-induced dynamics of $H = 2$ hopfions. The $H=2$ hopfions can be regarded as a composite structure obtained through the fusion of horizontally arranged two hopfions with different helicities by $\pi$~\cite{Ward2000,Hietarinta2012,Kasai2025}. As clarified in Sec.~\ref{sec:H1}, the SOT-driven dynamics of the $H = 1$ hopfion depends on its helicity, and particularly a change in the helicity by $\pi$ reverses the directions of the translational motions, as schematically illustrated in Fig.~\ref{fig_separation}(a). Thus, the SOT is expected to act as an external tensile force on the $H=2$ hopfion, pulling it in two opposite directions, as indeed observed in Fig.~\ref{fig6:H2_tension}(b). In this picture, the hopfion splitting occurs when the external effective tension overcomes the topological protection arising from the linking of the hopfion, as schematically shown in Fig.~\ref{fig_separation}(b).

This picture also explains the reappearance of the stable region above the splitting region in Fig.~\ref{fig5:H2_diagram}(a), where the $H=2$ hopfion ceases to split. In this region, the hopfion is severely distorted due to the tilt of the background spins, and therefore, the external effective tension induced by the SOT on the hopfion $(H=2)$ does not act in a manner that would tear it apart. Consequently, the hopfion exhibits translational and precessional motions rather than splitting. This is demonstrated in the velocity plot of Fig.~\ref{fig5:H2_diagram}(b), where the nonzero velocities are observed only in the stable green region above the splitting blue region in Fig.~\ref{fig5:H2_diagram}(a). For these reasons, the phase diagram in Fig.~\ref{fig5:H2_diagram}(a) exhibits a reentrant nature along the $\zeta$ axis.

\subsubsection{Quantitative estimate}
\label{Quantitative estimate}
To conclude this section for the $H = 2$ hopfion dynamics, let us estimate the magnetic field and electric current density required for the hopfion splitting, as well as the characteristic timescale over which the splitting occurs. Results obtained by solving Eq.~\eqref{eq:LLG} are converted into physical quantities with appropriate dimensions using the relations:
\begin{align}
    \tau = \frac{J_1 \gamma_0}{g \mu_{\rm B} \mu_0} t, \quad \zeta = \frac{\mu_{\rm B} \hbar P}{de J_1 M_{\rm s}}j,
    \label{eq:unitconversion}
\end{align}
where $J_1$ is the coupling constant in Eq.~\eqref{eq:model} with an energy dimension, $\gamma_0$ is the electron gyromagnetic ratio, $\mu_0$ is the vacuum permeability constant, $t$ is the real time, $\hbar$ is the Dirac constant, $P$ is the spin Hall angle, $d = N_z a$ is the thickness of the magnetic layer ($N_z = 101$ and $a$ are the number of site in the $z$ direction and the lattice constant), $e$ is the electron charge, $M_{\rm s} = \mu_{\rm B}/a^3$ is the saturation magnetization per site, and $j$ is the electric current density, respectively. To convert the dimensions of the results in Fig.~\ref{fig4:H2_separation}, we use $\gamma_0 = 2.211 \times 10^5~[\rm{mA}^{-1}\rm{s}^{-1}]$, $g = 2$, $\mu_{\rm{B}} = 5.788 \times 10^{-5}~[\rm{eV~T}^{-1}]$, $\mu_0 = 4\pi \times 10^{-7}~[\rm{NA}^{-2}]$, $\hbar = 1.055 \times 10^{-34}~[\rm{Js}]$, and $e =1.602 \times 10^{-19}~[\rm{C}]$. Following the previous study~\cite{Xichao2017}, we set $J_1 = 1~[\rm meV]$, $P = 0.4$, and $a = 4$~[\AA]. Under these settings, the dimensionless magnetic field $B = 0.003$ used in this section corresponds to $\tilde{B} = BJ_1/g \mu_{\rm B} = 26~[\rm mT]$. The electric current density with $\zeta = 0.002$ yields $j = 7.7 \times 10^{11}~[\rm Am^{-2}]$, which is comparable to the typical current density required for magnetization reversal by the SOT~\cite{Miron2011, Liu2012}. Assuming $\tau = 19000$ in Eq.~\eqref{eq:unitconversion} as splitting time from Fig.~\ref{fig4:H2_separation}, the hopfion splitting is expected to occur within $t = 12.5~[\rm ns]$. To conclude, the hopfion splitting occurs by several tens of nanoseconds under the magnetic field of $10^0 - 10^1$ mT and the electric current density of $10^{11} - 10^{12}~\rm Am^{-2}$, all of which are experimentally accessible with current technology.

\subsection{SOT-driven dynamics of hopfions with higher Hopf numbers}
\label{sec:H34}
Extending the analyses in the preceding sections, we investigate the dynamics of hopfions with $H=3$ in Sec.~\ref{sec:H3_separation} and $H=4$ in Sec.~\ref{sec:H4_separation}. As with the previous results for $H=2$, we demonstrate that an appropriate SOT also induces the splitting of hopfions with higher Hopf numbers, suggesting that the hopfion splitting by the SOT can occur universally.

\subsubsection{Case with $H=3$}
\label{sec:H3_separation}
\begin{figure}[tb]
  \centering
  \includegraphics[width=\hsize]{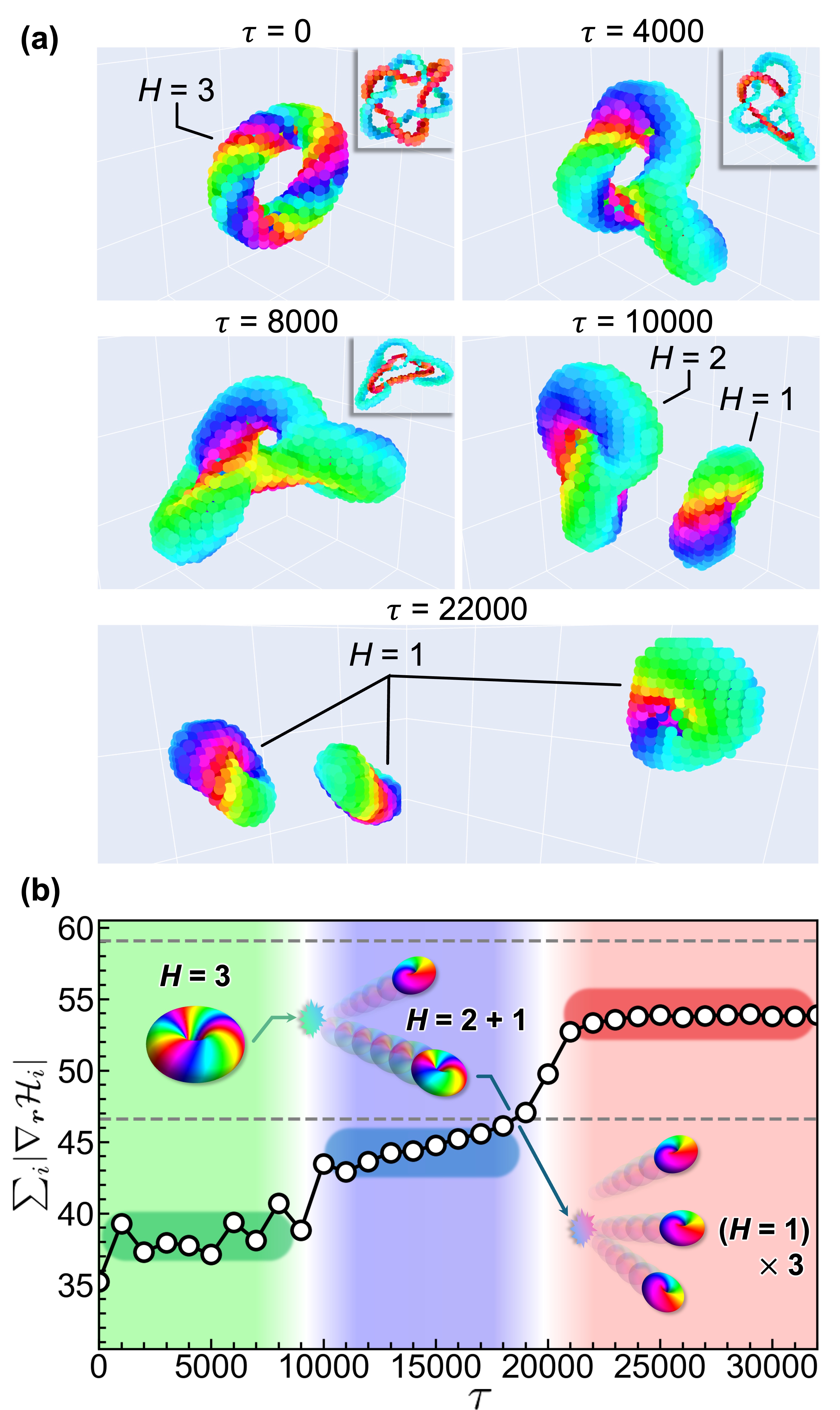}
  \caption{(a)~Snapshots of the hopfions with $H = 3$ and their preimages under $B = 0.003$ and $\zeta = 0.002$. The $H=3$ hopfion splits into $H=1$ and $H=2$ by $\tau=10000$, and eventually into three $H=1$ hopfions by $\tau=22000$. The viewpoint is varied with time for better visibility. (b)~Time evolution of the absolute sum of the spatial gradient of energy density during the dynamics in (a), with the schematic sequence of the hopfion splitting. Splitting occurs around the white time regions, during which the sum of the spatial gradient rapidly increases. The horizontal gray dashed line at $\sum_i |\nabla_{\bold r} \mathcal{H}_i| \simeq 59$ is the reference corresponding to three times the value of the gradient sum of a relaxed hopfion ($H=1$) in the absence of the SOT. Meanwhile, the line at $\sum_i |\nabla_{\bold r} \mathcal{H}_i| \simeq 47$ corresponds to the sum of the gradients of $H = 2$ and $H = 1$.}
  \label{fig7:H3_separation}
\end{figure}

Figure~\ref{fig7:H3_separation}(a) presents the dynamics of the $H=3$ hopfion with $B = 0.003$ and $\zeta = 0.002$. The initial state shown in the upper left panel of Fig.~\ref{fig7:H3_separation}(a) shows a slight elliptical deformation in its shape, consistent with the result in Ref.~\cite{Rybakov2022}. From the onset of the SOT at $\tau=0$ to around $\tau = 8000$, the hopfion shows complex deformations. In the snapshots at $\tau = 4000$ and $8000$, the spin structure appears to be composed of several independent hopfions that are barely bound together at the junctions. This result can be regarded as a competition with the SOT separating the hopfions and the effective tension highlighted in the previous section. By $\tau  = 10000$, the SOT overwhelms the binding of hopfions and it asymmetrically splits the hopfion with $H=3$ into $H = 2$ and $H = 1$. Afterward, the hopfion with $H=2$ further splits into two $H=1$ as shown in the snapshot by $\tau = 22000$.

Accompanying this two-step splitting process, the time evolution of the sum of energy gradients, as shown in Fig.~\ref{fig7:H3_separation}(b), exhibits two surges, with three plateaus corresponding to the three states: green for the single $H = 3$ hopfion, blue for the coexistence with $H=1$ and $H=2$, and red for the three $H=1$ hopfions. The two gray dashed lines at $\simeq 59$ and $47$ show the gradients corresponding to the sum of three independent $H=1$ hopfions and the sum of independent $H=2$ and $H=1$ hopfions. The gradient values in the intermediate and final stages of the splitting process are close to these values.

Changing $\zeta$, we find other steady states below and above the splitting region, similar to the $H=2$ case. One is the unstable phase for large $\zeta$, while the other two are stable phases --- one in which the hopfion remain stationary, and in the other in which it exhibits both translational and precession motions. The phase diagram for these dynamics will be shown in Fig.~\ref{fig10:hierarchy}, together with the comparison to other cases.

\subsubsection{Case with $H=4$}
\label{sec:H4_separation}
\begin{figure}[tb]
  \centering
  \includegraphics[width=\hsize]{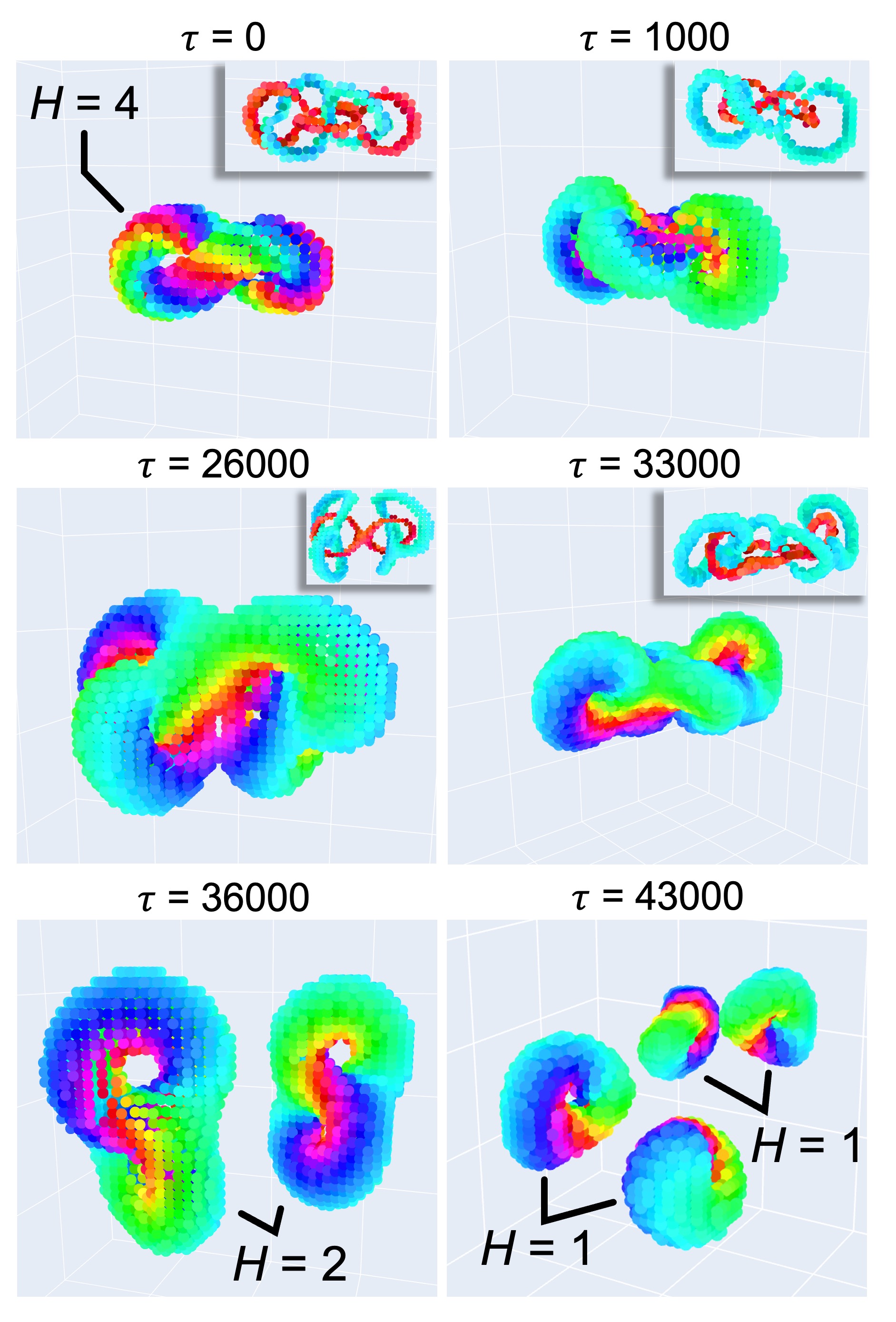}
  \caption{Snapshots of the hopfions with $H = 4$ and their preimages under $B = 0.003$ and $\zeta = 0.0021$. The $H=4$ hopfion splits into two $H=2$ hopfions by $\tau=36000$, and eventually into four $H=1$ hopfions by $\tau=43000$. The viewpoint is varied with time for better visibility.}
  \label{fig8:H4_separation}
\end{figure}

\begin{figure}[tb]
  \centering
  \includegraphics[width=\hsize]{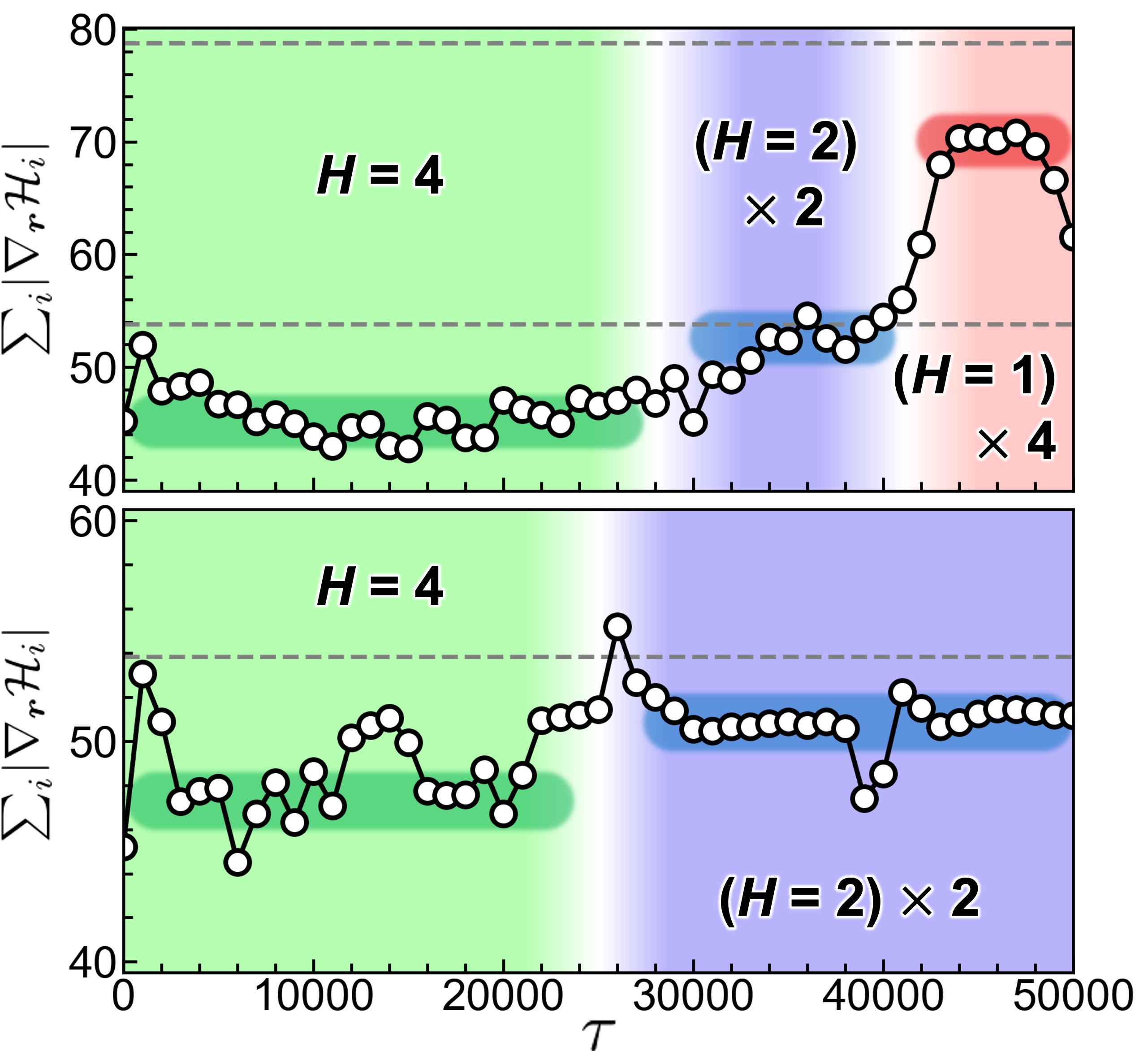}
  \caption{Time evolution of the absolute sum of the spatial gradient of energy density during the dynamics with $B = 0.003$ and $\zeta = 0.0021$ (upper panel) and $\zeta = 0.00235$ (lower panel). Splitting occurs around the white time regions, during which the sum of the spatial gradient rapidly increases. The horizontal gray dashed line at $\sum_i |\nabla_{\bold r} \mathcal{H}_i| \simeq 79$ in the upper panel is the reference corresponding to four times the value of the gradient sum of a relaxed hopfion ($H=1$) in the absence of the SOT. Meanwhile, the lines at $\sum_i |\nabla_{\bold r} \mathcal{H}_i| \simeq 54$ in both panels correspond to twice the value of the gradients of $H = 2$.}
  \label{fig9:H4_tension}
\end{figure}

Figure~\ref{fig8:H4_separation} displays snapshots of the $H = 4$ dynamics with $B = 0.003$ and $\zeta = 0.0021$. The initial state in the upper left panel of Fig.~\ref{fig8:H4_separation} has an ``8"-shaped isosurface on which $S^z \simeq 0$, consistent with that in Ref.~\cite{Rybakov2022}. The preimages in the inset appear to have more complex knots than those for the previous lower $H$ cases, and they are linked four times with each other. At $\tau = 1000$, we observe asymmetric deformations for both red and blue preimages. At $\tau = 26000$, the structure looks like a bound structure with two $H=2$ hopfions, each similar to the one in the $H=2$ dynamics at $\tau = 2000$ in Fig.~\ref{fig4:H2_separation}. By around $\tau = 36000$, the hopfion splits into two $H = 2$, which eventually split into four $H = 1$ hopfions, as shown in the snapshot at $\tau = 43000$. This two-step splitting dynamics is observed in the region of $0.0021 \lesssim \zeta \lesssim 0.0023$ for $B = 0.003$, while no splitting occurs for smaller $\zeta$, similar to the results for $H=2$ and $H=3$ in the previous sections.

Interestingly, for larger $\zeta$, in the range of $0.0023 \lesssim \zeta \lesssim 0.0024$, we find that the $H=4$ hopfions tend to split down to two $H=2$ hopfions, but each does not split into $H=1$ within the simulation timescale, $\tau < 50000$. This is understood by the comparison with the phase diagrams for $H=1$ and $H=2$ cases as follows. This region of $\zeta$ corresponds to the stable region above the splitting region for $H=2$ in Fig.~\ref{fig5:H2_diagram}(a), while it corresponds to the unstable region for $H=1$ in Fig.~\ref{fig3:H1_diagram}(a). Therefore, in this region, the $H = 4$ hopfion is not allowed to split into four independent $H=1$ hopfions. In contrast, the smaller $\zeta$ region of $0.0021 \lesssim \zeta \lesssim 0.0023$ corresponds to the splitting region for $H=2$, and therefore, the $H=2$ hopfions appearing in the intermediate time scale are able to split into $H=1$ hopfions.

Figure~\ref{fig9:H4_tension} shows the time evolution of the spatial gradients of energy for these two types of splitting dynamics. As shown in both the upper panel with $\zeta = 0.0021$ and the lower panel with $\zeta = 0.00235$, the energy gradient rapidly increases at $\tau \simeq 30000$, where the $H = 4$ hopfion splits into two $H = 2$ hopfions. Subsequently, since the $H=2$ can split into two $H=1$ hopfions under the SOT with $\zeta = 0.0021$, an additional increase appears at $\tau \simeq 42000$ in the upper panel. In contrast, such an increase is not observed in the lower panel, as the $H=2$ is stable for $\zeta = 0.00235$. The gray dashed lines at $\simeq 54$ and $79$ show the gradients corresponding to the sum of two independent $H=2$ hopfions and the sum of four independent $H=1$ hopfions, respectively. These values are comparable to those of the plateaus found in the time evolution~\cite{note_upturn}.

Applying an even larger $\zeta$, the hopfion once enters a region where it remains topologically stable as $H=4$, with showing translational and rotational motions, and eventually ends up in the unstable phase, like in the other cases. The details will be summarized as the steady-state phase diagram in Fig.~\ref{fig10:hierarchy} together with the other cases.

\subsection{Hierarchy in the steady-state phase diagrams under SOT}
\label{sec:hierarchy}
Building on the nonequilibrium dynamics revealed thus far, we explore the hierarchical relationships among different Hopf numbers. The hierarchy allows us to predict the SOT-driven dynamics of even higher-order hopfions.

To clarify the hierarchy, it is required to identify the topological properties of the nonequilibrium steady states obtained by long-time simulations. For this purpose, we calculate, in addition to the total Hopf number in Eq.~\eqref{eq:hopfnum} integrated over the entire system, the local Hopf number $\tilde{H}$ within a restricted region in space. Note that Eq.~\eqref{eq:hopfnum} yields the same value regardless of whether splitting occurs, as the sum of topological numbers is invariant, as long as the hopfions are preserved and not relaxed to a ferromagnetic state. The local integration around each hopfion can reveal whether and how splitting has occurred. Specifically, we consider a cube of appropriate size centered at the site where the integrand in Eq.~\eqref{eq:hopfnum} takes maximum. For instance, using this scheme for a hopfion with $H=2$, when it does not undergo splitting, the local Hopf number within the cube yields $\tilde{H} \simeq 2$, equal to the total one $H=2$. In contrast, when the hopfion is split into two $H = 1$ hopfions, the cubic region for computing $\tilde{H}$ encompasses only one of two $H = 1$ hopfions, resulting in $\tilde{H} \simeq 1$.

\begin{figure}[tb]
  \centering
  \includegraphics[width=\hsize]{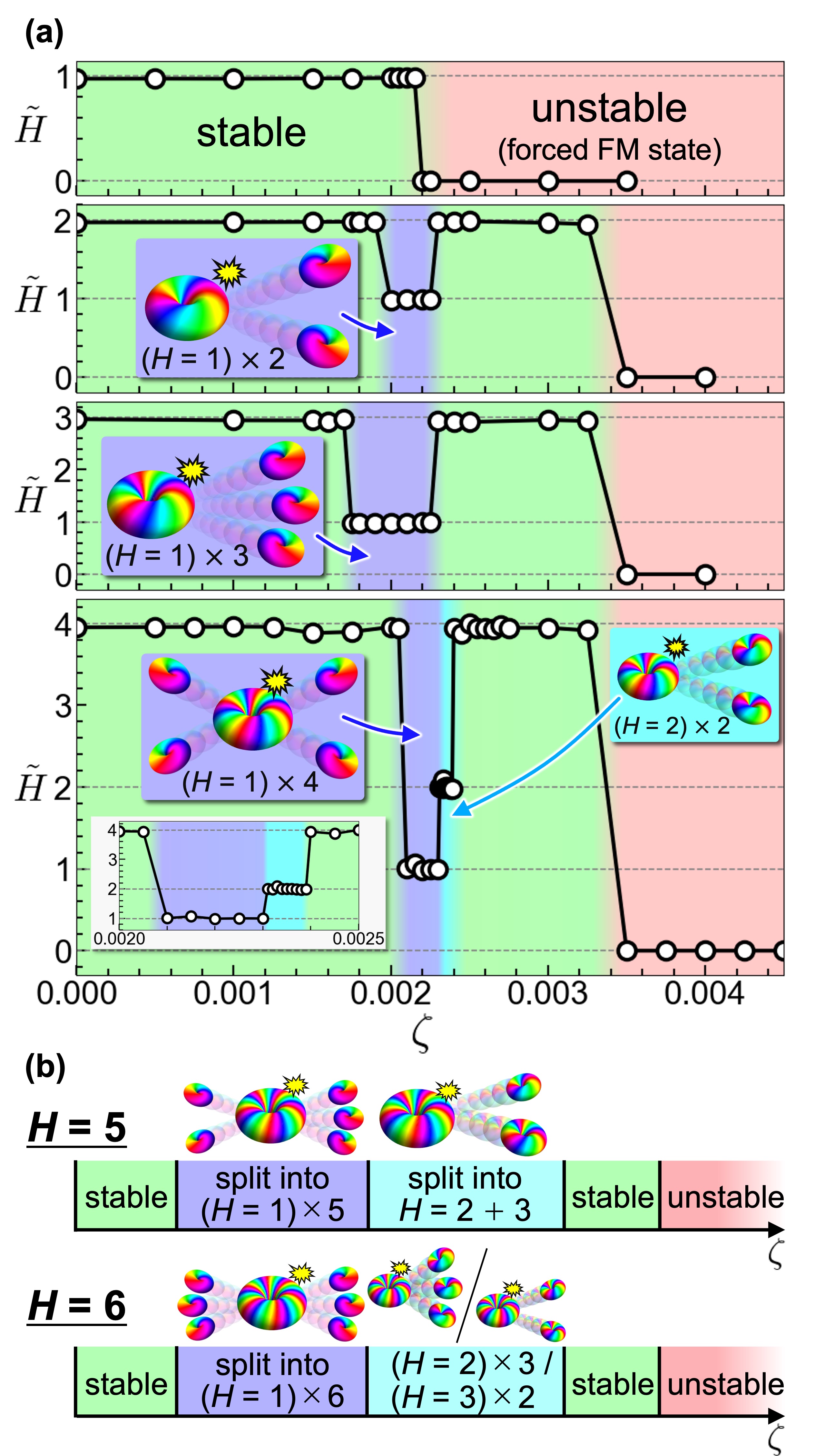}
  \caption{SOT-driven steady-state phase diagrams for different Hopf numbers while changing $\zeta$. We take $B=0.003$. (a)~The phase diagrams for $H = 1, 2, 3$, and $4$, with plots of the local Hopf number $\tilde{H}$, based on the simulation performed in Secs.~\ref{sec:H1}, \ref{sec:H2}, and \ref{sec:H34}. (b)~Predicted phase diagrams for $H = 5$ and $6$, based on the hierarchy observed in (a).}
  \label{fig10:hierarchy}
\end{figure}

Figure~\ref{fig10:hierarchy}(a) summarizes the steady-state phase diagrams at $B=0.003$ with the plots of the local Hopf numbers $\tilde{H}$ for the cases with $H=1, 2, 3$, and $4$. Note that $\tilde{H} \simeq H$ in the $H=1$ case. We find that some of the phase boundaries almost coincide with each other across the results for different Hopf numbers. Based on this observation, we demonstrate below that the SOT-driven dynamics of hopfions can be systematically understood.

For the SOT-induced dynamics of the $H = 2,3,$ and $4$ hopfions, the splitting down to $H = 1$ consistently begins around $\zeta \simeq 0.0018-0.0020$, as shown in Fig.~\ref{fig10:hierarchy}(a). The small differences in the critical values of $\zeta$ are not considered important, as the detailed dynamics is likely dependent on the initial conditions. This consistent onset suggests that the SOT-induced splitting of hopfions universally occur at a specific value of $\zeta$, determined by the strength of $B$, regardless of $H$. Another remarkable feature is that the hopfions become unstable consistently around $\zeta \simeq 0.0034$. As described in Eq.~\eqref{eq:forcedFM}, in this regime, the background magnetization lies entirely into the $xy$ plane, which induces severe deformation of the hopfion and eventually leads to its instability.

At $\zeta \simeq 0.0023$, the splitting phases of the $H = 2$ and $3$ hopfions switch again to the stable phases. Interestingly, the $H = 1$ hopfion enters the unstable phase at almost the same value of $\zeta$. This suggests that the stability of hopfions with a lower Hopf number under the application of the SOT determines whether the higher-$H$ hopfions exhibit splitting dynamics or not. For the $H=2$ hopfion, since the $H=1$ hopfion cannot exist stably for $\zeta \gtrsim 0.0023$, the splitting process halts and the spin structure remains in the $H=2$ state in this regime. Similarly, for the $H=3$ case, the hopfion does not exhibit splitting into $H=1$ and $H=2$ due to the instability of the $H=1$ hopfion in the same regime.

Interestingly, this behavior extends to the higher $H=4$ case. In this case, the $H=4$ hopfion no longer split into four $H=1$ in this regime. Instead, it splits into two $H=2$ within the cyan region of $0.0023 \lesssim \zeta \lesssim 0.0024$ in Fig.~\ref{fig10:hierarchy}(a). This switching of the splitting from $(H=1) \times 4$ to $(H=2) \times 2$ can be explained by the stability of $H=1$ and $H=2$ hopfions. As with the $H=2$ and $H=3$ cases discussed above, the instability of $H=1$ hopfion prevents splitting into four $H=1$, while the continued stability of the $H=2$ hopfion allows for splitting into two $H=2$. We note that in this regime the dynamics depends on the initial states; we plot representative results from simulations initiated with 15 distinct configurations, carefully examining the values of $\tilde{H}$ and corresponding snapshots. For larger $\zeta$, the splitting process halts, as in the lower $H$ cases.

These findings suggest a hierarchical structure in the steady-state phase diagrams for hopfions with different $H$: (i) the critical values of SOT for the stable--splitting transition (green to blue with increasing $\zeta$) and the stable--unstable transition (green to red with increasing $\zeta$) are nearly coincide across different $H$ for $H\geq 2$, and (ii) in the splitting regime, a hopfion can split into multiple hopfions with smaller $H$, provided that these resulting hopfions remain stable within the corresponding parameter region. Based on these rules, we can make some conjectures about the SOT-driven dynamics of hopfions with higher Hopf numbers. For instance, as shown in the upper panel of Fig.~\ref{fig10:hierarchy}(b), the hopfion with $H=5$ is expected to begin splitting into five $H=1$ hopfions around $\zeta \simeq 0.0018-0.0020$. As $\zeta$ increases, since $H=1$ becomes less favorable around $\zeta \simeq 0.0023$ while $H=2$ and $H=3$ remain stable, $H = 5$ might split into isolated hopfions with $H=2$ and $H=3$. With further increases in $\zeta$, the hopfion with $H=5$ may cease to split, and eventually becomes unstable for $\zeta \gtrsim 0.0034$, like all the previous results in Fig.~\ref{fig10:hierarchy}(a). By reasoning similarly, we can infer that a hopfion with $H=6$ would exhibit, at least, three possible splitting phases, as shown in the lower panel of Fig.~\ref{fig10:hierarchy}(b): the first being splitting into six hopfions with $H=1$, the second into three $H=2$ or two $H=3$. Our hierarchy rules cannot tell which splitting --  or whether both -- occurs in the latter process. As exemplified above, our findings enable the prediction of hierarchical structures in the steady-state phase diagrams for higher-order hopfions, which would be useful to tune the values of $B$ and $\zeta$ for controlling hopfions with general $H$.

\subsection{Splitting and recombination by time-dependent SOT}
\label{sec:recombination}
\begin{figure}[t!]
  \centering
  \includegraphics[width=\hsize]{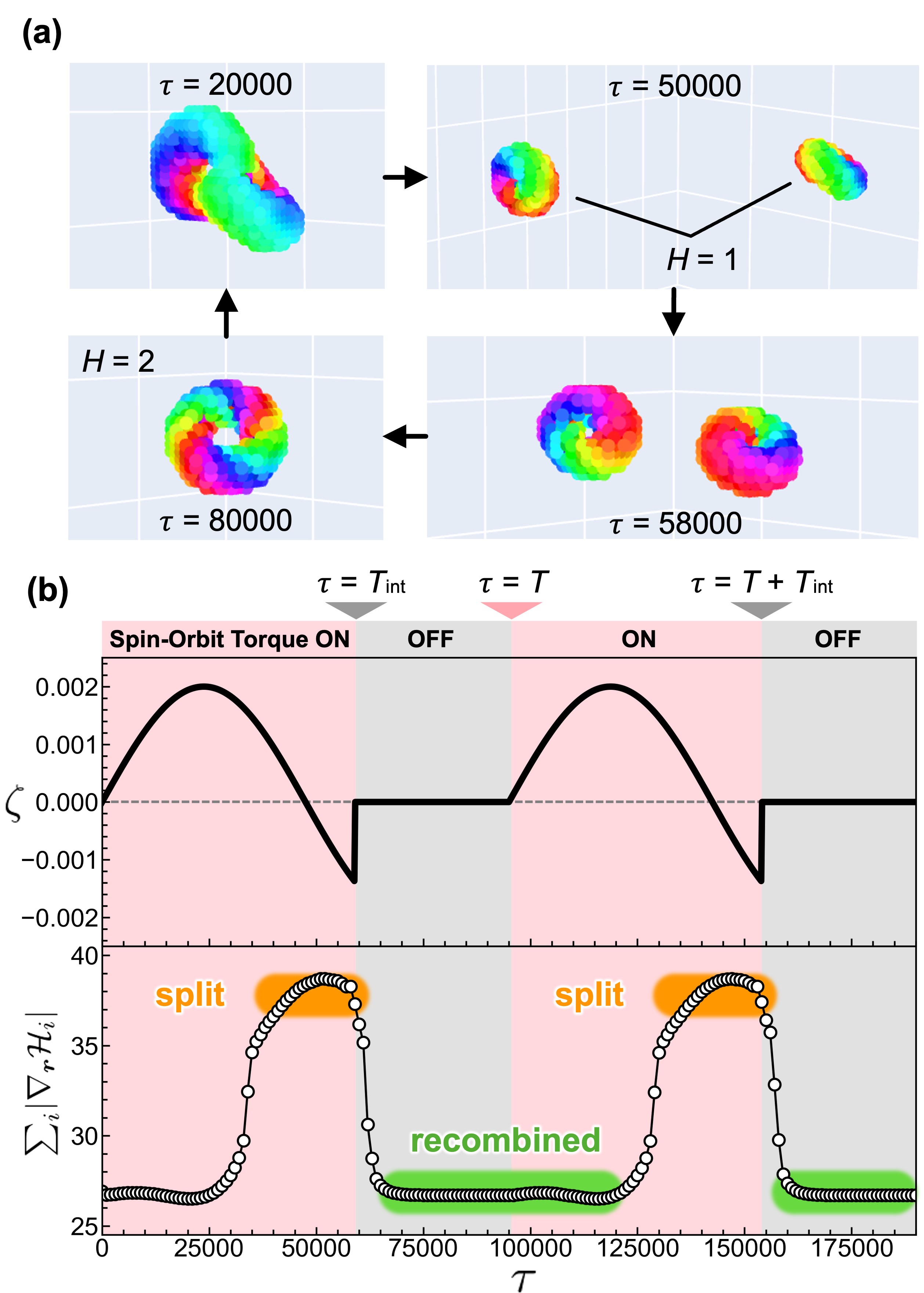}
  \caption{(a)~Snapshots in the splitting and recombination dynamics of the hopfion with $H = 2$ under a SOT with time dependence shown in the upper panel of (b) [Eq.~\eqref{eq:sinusoidal}]. The lower panel of (b) shows the norm of the spatial gradient of energy density introduced in Sec.~\ref{sec:effective_tension}. By toggling the sinusoidal SOT on and off, the splitting into two $H=1$ and their recombination can be repeatedly realized.}
  \label{fig11:sinusoidal}
\end{figure}

Finally, we demonstrate that the split hopfions can be recombined by designing the time dependence of the SOT. Our results in the previous sections showed that a higher-$H$ hopfion can be split into multiple lower-$H$ hopfions under the SOT. In addition, in the equilibrium state in the absence of SOT, hopfions experience a long-range attractive interaction, suggesting the possibility of their recombination into a higher-$H$ hopfion~\cite{Kasai2025}. By integrating these insights, we here demonstrate the splitting and recombination of an $H=2$ hopfion through careful tuning the SOT time profile. Specifically, we apply a time-dependent SOT, illustrated in the upper panel of Fig.~\ref{fig11:sinusoidal}(b), which is given by a sinusoidal waveform with periodic interruptions as
\begin{align}
    \zeta(\tau) =
    \begin{cases}
        ~\zeta_{\rm s} \sin\left(\frac{2\pi \tau}{T}\right)&(nT \leq \tau \leq  nT + T_{\rm int}) \\
        ~0&(nT + T_{\rm int} \leq \tau \leq  (n+1)T),
    \label{eq:sinusoidal}
    \end{cases}
\end{align}
with $n=0$, $1$, $2 \cdots$. We take $\zeta_{\rm s} = 0.002$, $T_{\rm int}=59000$, and the sinusoidal period $T = 95000$.

Figure~\ref{fig11:sinusoidal}(a) presents snapshots during the dynamics in the first period under the time-dependent SOT in Eq.~\eqref{eq:sinusoidal}. Since the sign of the SOT is positive up to $\tau=T/2$, its dynamics is similar to that in Fig.~\ref{fig4:H2_separation}, resulting in the splitting, as shown in the upper left panel ($\tau=20000$) to the upper right panel ($\tau=50000$) of Fig.~\ref{fig11:sinusoidal}(a). This splitting is also confirmed by the spatial gradient of energy plotted in the lower panel of Fig.~\ref{fig11:sinusoidal}(b). We note that the splitting dynamics under the sinusoidal SOT is sensitive to the period $T$; for example, for a shorter period $T=80000$, splitting does not occur. When the sign of the SOT switches to negative at $\tau=T/2$, each hopfion changes its direction of the translational motion. Consequently, as shown in the snapshot at $\tau=58000$, the hopfions approach each other again. After the interruption of the SOT at $\tau=59000$, the hopfions further approach and recombine due to the effective attractive interaction revealed in the previous study, as shown in the lower-left panel of Fig.~\ref{fig11:sinusoidal}(a) at $\tau=80000$~\cite{Kasai2025}. After a sufficient relaxation of the hopfion with $H = 2$ up to $\tau=95000$, the sinusoidal SOT is applied again, leading to the repeated splitting and recombination of the hopfions. We note that because the dynamics of the hopfions under negative $\zeta$ does not trace the trajectory for positive $\zeta$, the recombination demonstated in Fig.~\ref{fig11:sinusoidal} requires careful tuning of the sinusoidal period $T$ and the timing of the SOT interruption $T_{\rm int}$.

\section{Summary}
\label{sec:summary}
To summarize, we have investigated the nonequilibrium dynamics of hopfions with various Hopf numbers driven by the SOT, varying the magnetic field $B$ and the SOT strength $\zeta$. First, for an $H=1$ hopfion, we clarified that the hopfion exhibits both translational and precessional motions under SOT, dependent on their initial orientation and helicity. It is noteworthy that a $\pi$ change in its helicity, that is a $\pi$ rotation of the hopfion around its central axis, reverses the transverse components of both the velocity and precessional axis vectors characterizing their motions. Next, we performed similar analyses for hopfions with $H = 2$, and found that the SOT serves as an effective tension that promotes their forced splitting into two hopfions with $H = 1$ each. This splitting process can be induced on the nanosecond timescale using experimentally accessible current densities. The preimages before and after the splitting dynamics clearly illustrate the change in their linking configuration, demonstrating the high controllability of the hopfion’s knot topology with the SOT. Extending this analysis to the cases of $H = 3$ and $4$, we found that the splitting dynamics of the hopfions occurs universally, irrespective of the Hopf number. Interestingly, steady-state phase diagrams under the SOT reveal that the critical SOT values separating different steady-state behaviors are nearly coincide across different $H$. From the systematic analysis, we clarified the hierarchy in the steady-state phase diagrams for different $H$, indicating that the splitting of higher-$H$ hopfions are dictated by the stability of lower-$H$ hopfions under the SOT. These insights allow us to make conjectures on the SOT-driven dynamics of hopfions with general $H$. Finally, we demonstrated that a designed time-dependent SOT can repeatedly induce the splitting and recombination dynamics of an $H = 2$ hopfion.

Our discovery of the helicity-dependent SOT dynamics of $H = 1$ hopfions reveals a phenomenon not observed in the previous electric current-driven dynamics, offering a more flexible means of manipulating hopfions. Moreover, the universal splitting dynamics observed in higher-$H$ hopfions demonstrate the rich tunability of their linking topology. Assignment of each topological property controlable by the SOT to a distinct information state paves the way for the realization of multilevel and noise-tolerant magnetic memory devices. Since our simulation setup is based on conventional spin-torque switching protocols, our findings can be readily tested in experiments once hopfions are successfully generated in frustrate spin systems.


\section*{\label{acknowledge}Acknowledgments}
We thank M. Ezawa, Y. Kato, and K. Shimizu for fruitful discussions. This work was supported by the JSPS KAKENHI (No.~JP22K13998, JP23K25816, and JP25H01247) and JST PRESTO (No.~JPMJPR2595). S. K. was supported by the Program for Leading Graduate Schools (MERIT-WINGS) and JST SPRING, Grant Number JPMJSP2108. The computation in this work has been done using the facilities of the Supercomputer Center, the Institute for Solid State Physics, The University of Tokyo.


\bibliography{bibliography}

\end{document}